\documentclass[11pt,article]{elsarticle}

\usepackage{lineno,hyperref}
\usepackage[lmargin=.75in,rmargin=.75in,tmargin=1.in,bmargin=1in]{geometry}
\usepackage{amsmath}
\usepackage{amsthm}
\usepackage{amsfonts}
\usepackage{amssymb}
\usepackage{graphicx}
\usepackage{multicol}
\usepackage{wrapfig}
\usepackage{setspace}
\usepackage{epstopdf}
\usepackage{graphicx}
\usepackage{caption}
\usepackage{subcaption}
\usepackage{algorithm}
\usepackage[noend]{algpseudocode}
\newtheorem*{remark}{Remark}

\journal{arXiv}

\begin{document}

\begin{frontmatter}

\title{Bayesian inference and model comparison for metallic fatigue data}

\author[ICES]{Ivo Babu\v {s}ka}
\ead{babuska@ices.utexas.edu}
\author[KAUST]{Zaid Sawlan \corref{corrauthor}}
\cortext[corrauthor]{Corresponding author}
\ead{zaid.sawlan@kaust.edu.sa}
\author[KAUST,Ur]{Marco Scavino}
\ead{marco.scavino@kaust.edu.sa}
\author[St]{Barna Szab\'o}
\ead{szabo@wustl.edu}
\author[KAUST]{Ra\'ul Tempone}
\ead{raul.tempone@kaust.edu.sa}

\address[ICES]{ICES, The University of Texas at Austin, Austin, USA}
\address[KAUST]{CEMSE, King Abdullah University of Science and Technology, Thuwal, Saudi Arabia}
\address[St]{Washington University in St. Louis, St. Louis, USA}
\address[Ur]{Instituto de Estad\'{\i}stica (IESTA), Universidad de la Rep\'ublica, Montevideo, Uruguay}

\begin{abstract}
In this work, we present a statistical treatment of stress-life (S-N) data drawn from a collection of records of fatigue experiments that were performed on 75S-T6 aluminum alloys. Our main objective is to predict the fatigue life of materials by providing a systematic approach to model calibration, model selection and model ranking with reference to S-N data. To this purpose, we consider fatigue-limit models and random fatigue-limit models that are specially designed to allow the treatment of the run-outs (right-censored data). We first fit the models to the data by maximum likelihood methods and estimate the quantiles of the life distribution of the alloy specimen. To assess the robustness of the estimation of the quantile functions, we obtain bootstrap confidence bands by stratified resampling with respect to the cycle ratio. We then compare and rank the models by classical measures of fit based on information criteria. We also consider a Bayesian approach that provides, under the prior distribution of the model parameters selected by the user, their simulation-based posterior distributions. We implement and apply Bayesian model comparison methods, such as Bayes factor ranking and predictive information criteria based on cross-validation techniques under various a priori scenarios.
\end{abstract}


\end{frontmatter}

%\linenumbers

\section{Introduction}

Mechanical and structural components subjected to cyclic loading are susceptible to cumulative damage and eventual failure through an irreversible process called metal fatigue. Prediction of such fatigue through the expected service life of mechanical parts and assemblies is an important objective of numerical simulations used in mechanical and structural engineering practice. Based on such predictions, inspection intervals can be established. The frequency of these inspection intervals bears on the safety and costs of operation \cite{schijve, schijvebook, fatemi}.

%Furthermore, the expected service life of the mechanical parts depends on their material properties.

The fatigue characteristics of materials are established through fatigue tests performed on coupons, also called dogbone specimens, made of round bars or flat plates. The coupons are designed such that the stress is highest in the gauge section and that it remains substantially constant when the coupon is loaded in the axial direction.  In bending and torsion tests, the stress varies linearly over the cross section and is constant in the axial direction, for any fixed point in the cross section.

The number of cycles to failure, the peak stress and the cycle ratio are recorded for each experiment. The cycle ratio is defined as the minimum stress to maximum stress ratio. When an experiment is stopped before the specimen fails, then the test record is marked as a run-out. In some experiments, the specimen may buckle or fail outside of the gauge section. Such experiments are disregarded. State-of-the-art reviews on mechanical fatigue are presented in \cite{schijve} and \cite{fatemi}. Here, we focus on high-cycle (stress-life) fatigue.

The set of data pairs $(S_i , N_i)$, where $S_i$ is the stress and $N_i$  is the corresponding number of cycles at failure in the $i$th test, exhibits substantial statistical dispersion.  Interpretation and generalization of test data are essential for making risk-informed design decisions. Various statistical models such as lognormal, extreme value, Weibull and Birnbaum-Saunders distributions have been used for this purpose.

We consider different types of models that contain fatigue limit parameters. Although such models have been widely used (see, for example, \cite{faa, pasmee, rflm, ryan}), there is an ongoing debate concerning the existence of the fatigue limit \cite{pyttel, bathias}. Some authors use the terms ``endurance limit" or ``fatigue strength" instead of ``fatigue limit" \cite{schijve, rflm}. We distinguish between the fatigue limit, which is a physical notion, and the fatigue limit parameter, which is an unknown parameter, expressed in the same scale as the equivalent stress and calibrated for different models. Usually, data support curve fitting up to a certain number of cycles to failure only. Extrapolation beyond that number substantially increases uncertainty. For example, aluminum does not have a fatigue limit, since it will always fail if tested to a sufficient number of cycles. Therefore, the fatigue limit (fatigue strength)
of aluminum is reported as the stress level at which the material can survive after a large number of cycles. For the purposes of this paper, the number of cycles can be fixed at $2 \times 10^7$, since the available data do not contain substantially larger cycle values.

We employ a classical (likelihood-based) approach to fit and compare the proposed models using the 75S-T6 aluminum sheet specimen data set described in Section 2. Ultimately, we provide an analog Bayesian approach to fit and compare the models. Although Ryan used a Bayesian approach to find an optimal design for the random fatigue-limit model \cite{ryan}, we are, to our knowledge, the first to use Bayesian methods to analyze and compare fatigue models.

The remainder of this paper is organized as follows. Section \ref{sec2} introduces the main characteristics of the fatigue tests conducted at the Battelle Memorial Institute on 85 75S-T6 aluminum sheet specimens by means of a Krouse direct repeated-stress testing machine. The data set with the fatigue test results is available as a csv file in the supplemental material to this paper. This data set contains run-outs. Section \ref{sec3} presents classical statistical models of fatigue test results. In Subsection \ref{sec3a}, we first consider a classical statistical fitting technique, called logarithmic fit, for illustration purposes only, that does not take in to account the presence of run-outs. Subsequently, we introduce fatigue-limit models and random fatigue-limit models, which are both specially designed to fit data in the presence of run-outs. We fit two fatigue-limit models, whose mean value function is same as in the logarithmic fit, with constant and non-constant variance functions, by constructing the corresponding likelihood functions and estimating all the unknown parameters that define the S-N curves by means of the maximum likelihood method. The fatigue limit parameter assessment under both models can be done by computing numerically tailored functions from their joint likelihoods, usually called profile likelihoods \cite{pawitan}. Later, we extend these models by assuming that the fatigue limit parameter is a random variable. To clarify the fitting procedure that provides estimates for S-N curves and predictions of fatigue life, we consider two random fatigue-limit models and their extensions, where a non-constant variance function is used. The assessment of the fatigue limit parameter is then summarized by comparing the estimated probability density functions of the four fitted models. Subsection \ref{sec3b} includes the computation of bootstrap confidence bands for the S-N curves and bootstrap confidence intervals for the maximum likelihood estimates. Subsection \ref{sec3c} is dedicated to comparison of the models by some widely used information criteria. Section \ref{sec4} focuses on the Bayesian analysis of some of the models. In Subsection \ref{sec4a}, three of the models analyzed using the likelihood approach are embedded in a Bayesian framework that we characterize based on informative priors and on non-informative priors. We use Bayesian computational techniques to estimate the posterior probability density function of each individual parameter of the six fitted models as well as the bivariate posterior probability functions of all the combinations of two parameters out of the total number of parameters for any of the six fitted models. Subsection \ref{sec4b} presents the Bayesian model comparison approach, which includes the Bayes factor and predictive information criteria. The Bayes factor is approximated by means of the Laplace method and the Laplace-Metropolis method. The Bayes factor is used to evaluate the fit of Bayesian models while the predictive information criteria are used to compare models based on their predictive accuracy.

\section{The 75S-T6 aluminum sheet specimens data set}
\label{sec2}
Data are available from 85 fatigue experiments that applied constant amplitude cyclic loading to unnotched sheet specimens of 75S-T6 aluminum alloys \cite[table 3, pp.22--24]{gbj}. The following data are recorded for each specimen:
\begin{itemize}
\item the maximum stress, $S_{max}$, measured in ksi units.
\item the cycle ratio, $R$, defined as the minimum to maximum stress ratio.
\item the fatigue life, $N$ , defined as the number of load cycles at which fatigue failure occurred.
\item a binary variable (0/1) to denote whether or not the test had been stopped prior to the occurrence of failure (run-out).
\end{itemize}
In 12 of the 85 experiments, the specimens remained unbroken when the tests were stopped. The recorded number of load cycles for these 12 experiments is the lower bound of an interval in which failure would have occurred had the test been continued. If specimens buckled or failed outside the test section, they are not included in the data set.

\section{Classical approach}
\label{sec3}

\subsection{Model calibration}
\label{sec3a}

There are many linear and nonlinear models (S-N curves) that have been used to predict fatigue life, $N$, in terms of the stress, $S$. A good list of these models can be found in \cite{castillo}. In this section, we consider relevant nonlinear regression models used with the 75S-T6 data set. For the sake of completeness, we first show how the fitting procedure works for a model that does not take into account the run-out feature of some observations. This so-called ``equivalent stress equation model" was used in \cite{faa}. Secondly, we introduce some fatigue-limit models that are tailored to work well in the presence of run-out observations, similar to \cite{pasmee} and \cite{rflm}, and we calibrate each of these models by using the maximum likelihood method.

In all the proposed models, the quantities of interest are the prediction of fatigue life, given the test stress and the cycle ratio, and the estimation of the fatigue limit parameter. The fatigue life predictions are summarized by means of the quantile functions. We plot the median (S-N curve), the $0.95$ quantile and the $0.05$ quantile.

Prior to the fitting of any statistical model, the fatigue data obtained for particular cycle ratios need to be generalized to arbitrary cycle ratios. For this purpose, the equivalent stress, $S_{eq}$, is then defined as $S_{eq}^{(q)} = S_{max}\,(1-R)^{q}$, where $q$ is a fitting parameter. This definition is also used in \cite{faa} and \cite{walker}.

We first consider the logarithmic fit as defined in \cite{sza} and \cite{faa}; that is, 
\begin{equation}
\mu(S_{eq}^{(lg)}) = A_1 + A_2 \log_{10}( S_{eq}^{(lg)} - A_3),
\label{ansatz}
\end{equation}
using the objective function proposed in (\cite{sza}),
\begin{equation}
e_{\textrm{std}} = \left(\frac{\sum_{i=1}^{n} (\log_{10}(n_i) - \mu(S_{eq}^{(lg)}))^2}{n-p}\right)^{1/2},
\label{model0}
\end{equation}
where $n$ is the number of data points and $p$ is the number of fitting parameters (namely $A_1, A_2, A_3$ and $q$). \\
The resulting estimated mean value function is given by
\begin{equation*}
\mu(S_{eq}^{(lg)}) = 10.07 - 3.54 \,\log_{10}( S_{eq}^{(lg)} - 25.41)\,,
\end{equation*}
where $S_{eq}^{(lg)} = S_{max}\,(1-R)^{0.5147}$ and the value of the objective function is $e_{std} = 0.5195\,$. 

\begin{figure}[h!]
\centering
\includegraphics[width=18cm]{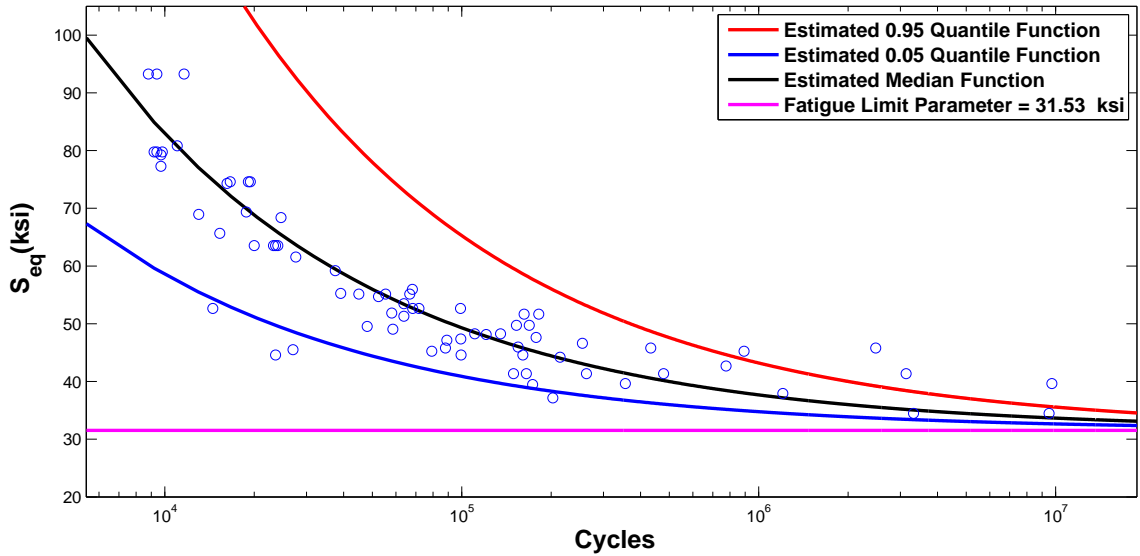}
\caption{Logarithmic fit of the 75S-T6 data set without run-outs. The fatigue life prediction increases toward infinity as the equivalent stress, $S_{eq}$, tends to the estimated fatigue limit parameter (horizontal asymptote) for any estimated quantile function. }
\label{fit0}
\end{figure}

\begin{remark}
Run-outs will introduce a bias error in the estimate when this approach is used. The resulting estimated mean value function, without the run-outs, is given by
\begin{equation*}
\mu(S_{eq}^{(lg)}) = 7.71 - 2.17 \,\log_{10}( S_{eq}^{(lg)} - 31.53)\,,
\end{equation*}
where $S_{eq}^{(lg)} = S_{max}\,(1-R)^{0.4633}$ and the value of the objective function is $e_{std} = 0.3673\,$.
Clearly, removing the run-outs increases the value of the fatigue limit. Figure \ref{fit0} shows the estimated quantile functions for the logarithmic fit, with the estimated fatigue limit parameter equal to $31.53$ ksi. We point out that the estimated fatigue limit is equal to $31.53/(2^{0.4633}) = 22.87$ ksi, since the fatigue limit is the value of the maximum stress when the cycle ratio, $R$, is equal to $-1$ (the ``fully reversed" condition).
\end{remark}

\subsubsection{Model Ia}
\label{model1} 
Let $A_3$ be the fatigue limit parameter. At each equivalent stress with $S_{eq} > A_3$, the fatigue life, $N$, is modeled by means of a lognormal distribution.
This implies that $\log_{10}(N)$ is modeled with a normal distribution with mean $\mu(S_{eq})$ and standard deviation $\sigma(S_{eq})$.
We generalize the logarithmic fit by assuming that
\begin{itemize}
\item $\mu(S_{eq}) = A_1 + A_2 \, \log_{10}( S_{eq} - A_3)\,,\:\:\textrm{if}\:\:S_{eq} > A_3$
\item $\sigma(S_{eq}) = \tau\,.$
\end{itemize}
Moreover, the model is now properly tailored to include the available censored fatigue data (run-outs). Given the sample data, $\bold{n} = (n_1, \ldots, n_m)$ and assuming that the observations are independent, the likelihood function is therefore given by
\begin{equation*}
L(A_1, A_2, A_3, \tau, q; \bold{n}) = \prod_{i=1}^{m} \left[ \frac{1}{n_i \log(10)} g(\log_{10}(n_i)\,;\mu(S_{eq})\,,\sigma(S_{eq})) \right]^{\delta_i}  \, 
\left[ 1- \Phi \left( \frac{\log_{10}(n_i) - \mu(S_{eq})}{\sigma(S_{eq})} \right) \right]^{1 - \delta_i} \,,
\end{equation*}
where $g(t; \mu, \sigma) = \frac{1}{\sqrt{2 \pi} \,\sigma} exp \left\{ - \frac{(t - \mu)^2}{2 \sigma^2} \right\}\,,$ $\Phi$ is the cumulative distribution function of the standard normal distribution, and
\begin{equation*}
\delta_i = \left\{
\begin{array}{rl}
1 & \text{if } n_i \text{ is a failure}\\
0 & \text{if } n_i \text{ is a run-out\,.}
\end{array} \right.
\end{equation*} 

\begin{figure}[h!]
\includegraphics[width=18cm]{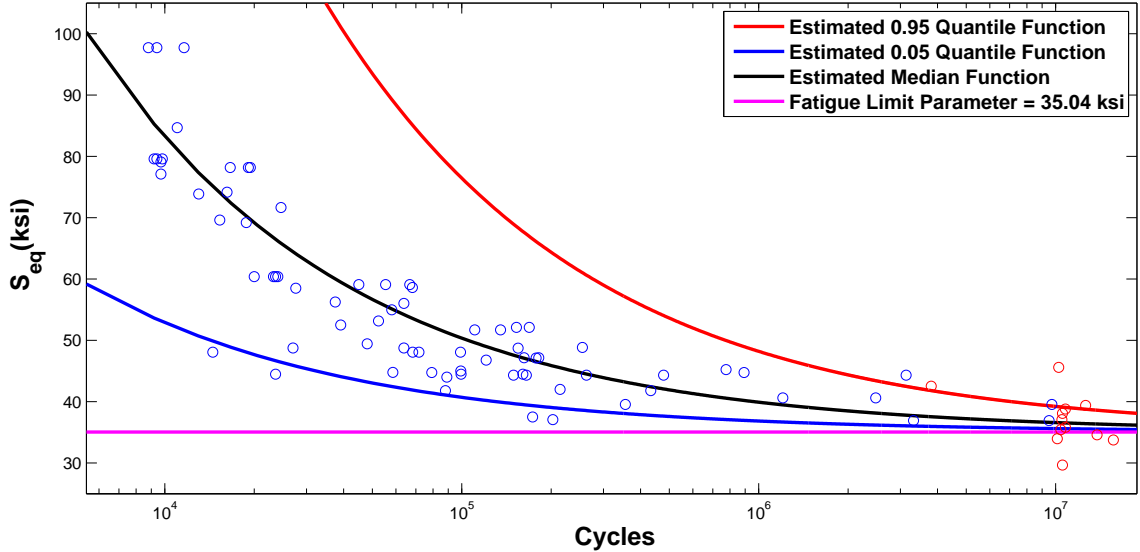}
\caption{Model Ia fit of the 75S-T6 data set. Under the assumption that $S_{eq}$ has constant variance, the addition of the run-outs (red circles) and fitting a model designed to handle right-censored data has the effect of enlarging the gap between the median and the $0.95$ quantile (in the upper range of values of $S_{eq}$, the number of cycles to attain failure has substantially increased with respect to the logarithmic fit) and between the median and the $0.05$ quantile (in the lower range of values of $S_{eq}$, the number of cycles to attain failure has decreased with respect to the logarithmic fit). The fatigue limit parameter estimate (purple line) is closer to the observed failures (blue circles) with smallest values of $S_{eq}$ than the same estimate using the logarithmic fit (Figure \ref{fit0}).}
\label{fit1}
\end{figure} 

This model is characterized by five parameters: $\theta = (A_1, A_2, A_3, q, \tau)$, whose maximum likelihood (ML) estimate, obtained by calibrating the model with the data, is
\begin{equation*}
\mu(S_{eq}) = 7.38 - 2.01 \, \log_{10}( S_{eq} - 35.04)\,,
\end{equation*}
where $S_{eq} = S_{max}\,(1-R)^{0.5628}$ and $\tau = 0.5274 \,.$ The maximum likelihood estimates are summarized in Table \ref{mle1a}. The corresponding fit is shown in Figure \ref{fit1} (blue circles = observed failures; red circles = run-outs). The difference between Model Ia and the logarithmic fit shows the importance of including the run-outs especially in the estimation of the fatigue limit. Run-outs that correspond to equivalent stress levels greater than the fatigue limit parameter are called significant run-outs. Only significant run-outs contribute to estimating the parameters. In this case, eight of the 12 run-outs were significant.

\begin{table}[h!]
\begin{center}
\caption{Maximum likelihood estimates for Model Ia}
\begin{tabular}{|c|c|c|c|c|c|c|}
\hline
 & $A_1$ & $A_2$ & $A_3$ & $q$ & $\tau$ \\
\hline
Model Ia & 7.38 &   -2.01 &  35.04 &  0.5628 &   0.5274  \\
\hline
\end{tabular}
\label{mle1a}
\end{center}
\end{table}

\subsubsection{Model Ib} 
\label{model1b}
We extend the model proposed in Subsection \ref{model1} by allowing a non-constant standard deviation as in \cite{pasmee}:
\begin{itemize}
\item $\mu(S_{eq}) = A_1 + A_2 \, \log_{10}( S_{eq} - A_3) \,,\:\:\textrm{if}\:\:S_{eq} > A_3$
\item $\sigma(S_{eq}) = 10^{(B_1 + B_2 \log_{10}(S_{eq}))} \,,\:\:\textrm{if}\:\:S_{eq} > A_3$
\end{itemize}

\begin{figure}[h!]
\centering
\includegraphics[width=18cm]{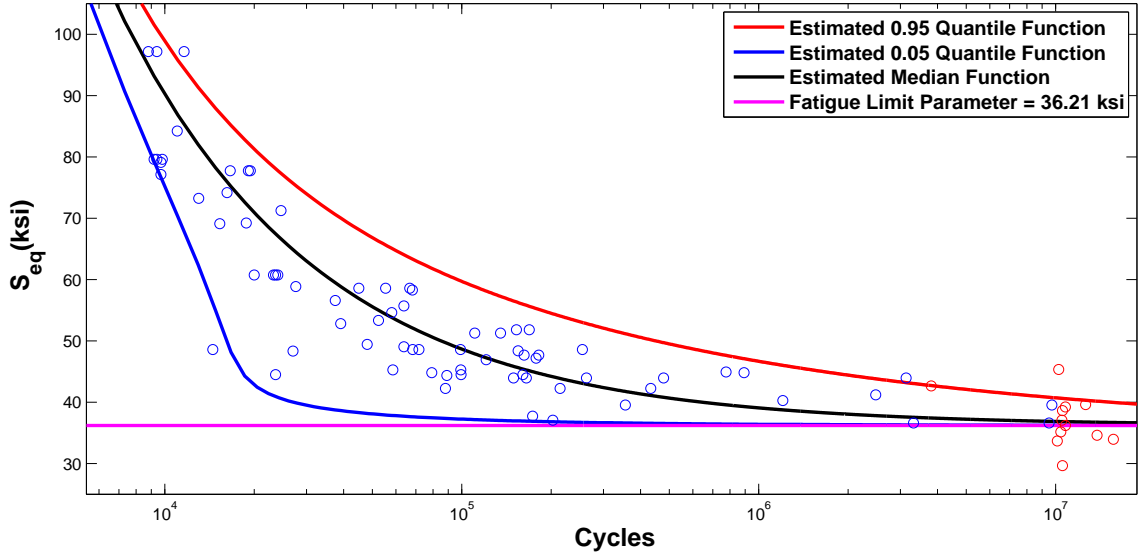}
\caption{Model Ib fit of the 75S-T6 data set. Allowing non-constant variance of $S_{eq}$ in a censored data model has the effect of reducing the gap between the median and both the $0.95$ and $0.05$ quantiles along the upper range of values of $S_{eq}$. In the case of the lower range of values of $S_{eq}$, the gap between the median and the 0.05 quantile has increased with respect to the Model Ia fit (Figure 2). The estimate of the fatigue limit parameter is very close to the minimum value of $S_{eq}$ that leads to failure. The estimated fatigue limit is $24.71$ ksi.}
\label{fit1b}
\end{figure} 

In this model, there are six parameters: $\theta = (A_1, A_2, A_3, q, B_1, B_2)$, and their ML estimates are
\begin{eqnarray*}
\mu(S_{eq}) = 6.72 - 1.57 \, \log_{10}( S_{eq} - 36.21), \\
\sigma(S_{eq}) = 10^{(4.55 - 2.89 \log_{10}(S_{eq}))},
\end{eqnarray*}
where $S_{eq} = S_{max}\,(1-R)^{0.5510}$. The maximum likelihood estimates are summarized in Table \ref{mle1b}. The corresponding fit is shown in Figure \ref{fit1b} (blue circles = observed failures; red circles = run-outs). Figure \ref{fit1b} shows that the uncertainty in predicting fatigue life decreases with high values of the equivalent stress when compared to Model Ia. However, the uncertainty increases for values of the equivalent stress that are close to the estimated fatigue limit parameter. In Model Ib, there are seven significant run-outs because the fatigue limit parameter has increased to $36.21$ ksi. When $A_3 < S_{eq} < 100$, the estimated standard deviation ranges between $1.11$ and $0.059\,,$ supporting the assumption of a non-constant standard deviation.

\begin{table}[h!]
\begin{center}
\caption{Maximum likelihood estimates for Model Ib}
\begin{tabular}{|c|c|c|c|c|c|c|c|}
\hline
& $A_1$ & $A_2$ & $A_3$ & $q$ & $B_1$ & $B_2$ \\
\hline
Model Ib & 6.72 &  -1.57  & 36.21 & 0.5510 &   4.55 &  -2.89  \\
\hline
\end{tabular}
\label{mle1b}
\end{center}
\end{table}

\begin{figure}[h!]
\centering
\includegraphics[width=10cm]{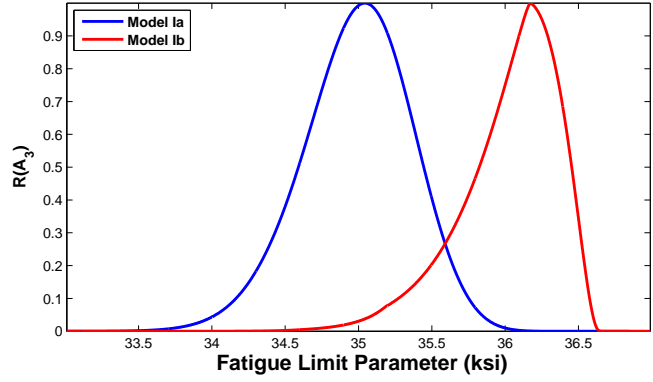}
\caption{Profile likelihood estimates for the fatigue limit parameter, $A_3$, with Model Ia fit (blue curve) and Model Ib fit (red curve). The two fitted fatigue-limit models display different ranges for the most plausible values of the fatigue limit parameter, $A_3$, a feature that is amplified by the left-skewed profile likelihood under Model Ib.}
\label{profile}
\end{figure}

\begin{remark}[\textbf{Profile likelihoods}]
To assess the plausibility of a range of values of the fatigue limit parameter, $A_3$, we construct the profile likelihood \cite[p. 294]{pasmee}:
\begin{equation}
R(A_3) = \max_{\theta_0} \left[ \frac{L(\theta_0,A_3)}{L(\hat{\theta})} \right] \,,
\end{equation}
where $\theta_0$ denotes all parameters except for the fatigue limit parameter, $A_3$, and $\hat{\theta}$ is the ML estimate of $\theta$.\\
Figure \ref{profile} shows the profile likelihood functions for $A_3$ corresponding to the models in Subsections \ref{model1} and \ref{model1b}. As in \cite{pasmee}, approximate $100(1-\alpha) \%$ confidence intervals for $A_3$ based on the calibrated profile likelihoods are given by: $\{ A_3 : -2 \log ( R(A_3) ) \leq \chi_{1;1-\alpha}^2 \}\,,$
where $\chi_{1;1-\alpha}^2$ is the $100(1-\alpha)$ percentile of a chi-square distribution with $1$ degree of freedom. The approximate $95 \%$ confidence intervals for $A_3$ are $(32.45, 36.28)$ and $(34.36, 36.88)$ for models Ia and Ib, respectively. We can see that each model suggests a different range for the fatigue limit parameter, $A_3$. We therefore need to systematically choose which model is better to assess the value of $A_3$.
\end{remark}

%Both models predict the fatigue limit within an interval of approximately $10\%$.

\subsubsection{Model IIa} 
\label{model2}
We now extend the model proposed in Subsection \ref{model1} to allow a random fatigue limit parameter as in \cite{rflm} :
\begin{itemize}
\item $\mu(S_{eq}) = A_1 + A_2 \, \log_{10}( S_{eq} - A_3)\,,\:\:\textrm{if}\:\:S_{eq} > A_3$.
\item $\sigma(S_{eq}) = \tau\,.$
\item $\log_{10}(A_3) \sim N(\mu_{f}, \sigma_{f})$.
\end{itemize}

Here, we assume that $\log_{10}(N)$ given $A_3 < S_{eq}$ is modeled with a normal distribution with mean $\mu(S_{eq})$ and standard deviation $\sigma(S_{eq})$. In this case, the probability density function (pdf) of $\log_{10}(N)$ is obtained by marginalizing $A_3$:
\begin{equation*}
f_{\log_{10}(N)} (u\,; \theta) = \int_{0}^{S_{eq}} h(u\,;\mu(S_{eq})\,,\sigma(S_{eq})) \, \ell(w\,;\mu_{f}\,,\sigma_{f})\, dw \,,
\end{equation*}
where $\theta = (A_1, A_2, \mu_{f}, \sigma_{f}, q, \tau)$, $h(u\,;\mu(S_{eq})\,,\sigma(S_{eq}))$ is the conditional density of $\log_{10}(N)$ given $A_3$, and $\ell(w\,;\mu_{f}\,,\sigma_{f})$ is the marginal density of $A_3$.
Similarly, the marginal cumulative distribution function (cdf) of $\log_{10}(N)$ is given by
\begin{equation*}
F_{\log_{10}(N)} (u \,; \theta) = \int_{0}^{S_{eq}} \Phi \left( \frac{u - \mu(S_{eq})}{\sigma(S_{eq})} \right) \, \ell(w\,;\mu_{f}\,,\sigma_{f})\, dw \,,
\end{equation*}
where $\Phi$ is the cumulative distribution function of the standard normal distribution. The functions $f_{\log_{10}(N)}$ and $F_{\log_{10}(N)}$ no longer have closed forms and must be numerically evaluated. Global adaptive quadrature is used to approximate the integrations (see \cite{quad}). 

Assuming independent observations, the likelihood function of $\theta = (A_1, A_2, \mu_{f}, \sigma_{f}, q, \tau)$ is therefore given by
\begin{equation}
\label{likefun}
L(\theta; \{\log_{10}(n_1), \ldots, \log_{10}(n_m)\}) = \prod_{i=1}^{m} \left[ f_{\log_{10}(N)} (\log_{10}(n_i)\,; \theta) \right]^{\delta_i}  \, \left[ 1- F_{\log_{10}(N)} (\log_{10}(n_i)\,; \theta) \right]^{1 - \delta_i} \,,
\end{equation}
where 
\begin{equation*}
\delta_i = \left\{
\begin{array}{rl}
1 & \text{if } n_i \text{ is a failure}\\
0 & \text{if } n_i \text{ is a run-out\,.}
\end{array} \right.
\end{equation*} 

\subsubsection{Model IIb} 
\label{model2b}
We can also consider a random fatigue-limit model with the smallest extreme value (sev) distribution as in \cite{rflm}:
\begin{itemize}
\item $\mu(S_{eq}) = A_1 + A_2 \, \log_{10}( S_{eq} - A_3)\,,\:\:\textrm{if}\:\:S_{eq} > A_3$.
\item $\sigma(S_{eq}) = \tau\,.$
\item the density of $\log_{10}(A_3)$ is $\phi (t; \mu_{f}, \sigma_{f})$.
\item the conditional density of $\log_{10}(N)$ given $A_3 < S_{eq}$ is $\phi (t; \mu(S_{eq}), \sigma(S_{eq}))\,,$
\end{itemize}
where $ \phi(t; \mu, \sigma) = \frac{1}{\sigma} exp \left\{ \left( \frac{t - \mu}{\sigma} \right) - exp \left( \frac{t - \mu}{\sigma} \right) \right\} $ is the sev probability density function with location parameter $\mu$ and scale parameter $\sigma$ \cite[Chapter 4]{meekerbook}. The likelihood function has the same form as in equation \eqref{likefun}. In other words, the conditional fatigue life, $N$, and the fatigue limit parameter, $A_3$, are modeled by a Weibull distribution. 

\begin{figure}[h!]
\centering
\includegraphics[width=18cm]{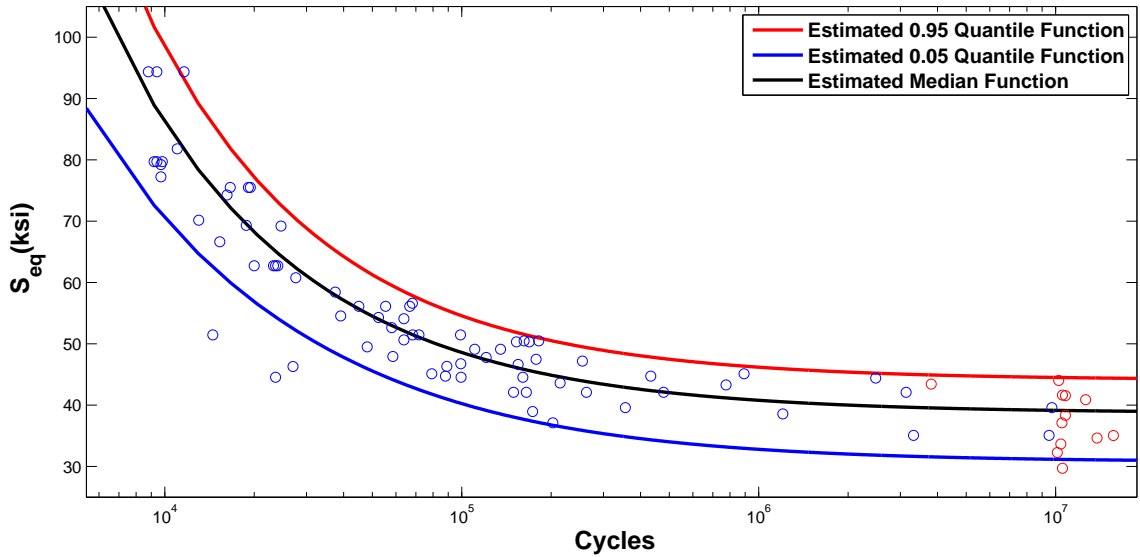}
\caption{Model IIb fit of the 75S-T6 data set. The fitting of a random fatigue-limit model for censored data has the effect that the estimated quantiles converge fast to an horizontal asymptote. Unlike fatigue-limit models, the random fatigue-limit model has the property that each estimated quantile approaches a different horizontal asymptote.}
\label{fit2b}
\end{figure} 

Table \ref{mle1} shows the maximum likelihood estimates and the maximum likelihood values obtained for Model IIa and Model IIb. The estimated parameters for both models are similar except for the parameters, $\sigma_f$ and $\tau$, which have smaller values with Model IIb. As a consequence, Model IIb has a smaller maximum likelihood value. Since models IIa and IIb have the same number of parameters, we can conclude that Model IIb is better than Model IIa. It is thus sufficient to present the corresponding fit of Model IIb (Figure \ref{fit2b}). 

\begin{table}[h!]
\begin{center}
\caption{Maximum likelihood estimates for Model IIa and Model IIb.}
\begin{tabular}{|c|c|c|c|c|c|c|c|c|}
\hline
  & $A_1$ & $A_2$ & $\mu_f$ & $\sigma_f$ & $q$ & $\tau$ & $\log(L^*)$\\
\hline
Model IIa & 6.53 & -1.51 & 1.58 & 0.0473 & 0.4888 & 0.1447 & -913.42 \\
\hline
Model IIb & 6.51 & -1.47 & 1.60 & 0.0385 & 0.4886 & 0.0852 & -907.31 \\
\hline
\end{tabular}
\label{mle1}
\end{center}
\end{table}

\subsubsection{Model IIc} 
\label{model2c}
The random fatigue-limit model proposed in Subsection \ref{model2} is extended by allowing non-constant standard deviation:
\begin{itemize}
\item $\sigma(S_{eq}) = 10^{(B_1 + B_2 \log_{10}(S_{eq}))}\,.$
%\,,\:\:\textrm{if}\:\:S_{eq} > A_3$
\end{itemize}

\subsubsection{Model IId} 
\label{model2d}
The random fatigue-limit model proposed in Subsection \ref{model2b} is extended by allowing non-constant standard deviation:
\begin{itemize}
\item $\sigma(S_{eq}) = 10^{(B_1 + B_2 \log_{10}(S_{eq}))}\,.$ 
%\,,\:\:\textrm{if}\:\:S_{eq} > A_3$
\end{itemize}

\begin{figure}[h!]
\centering
\includegraphics[width=18cm]{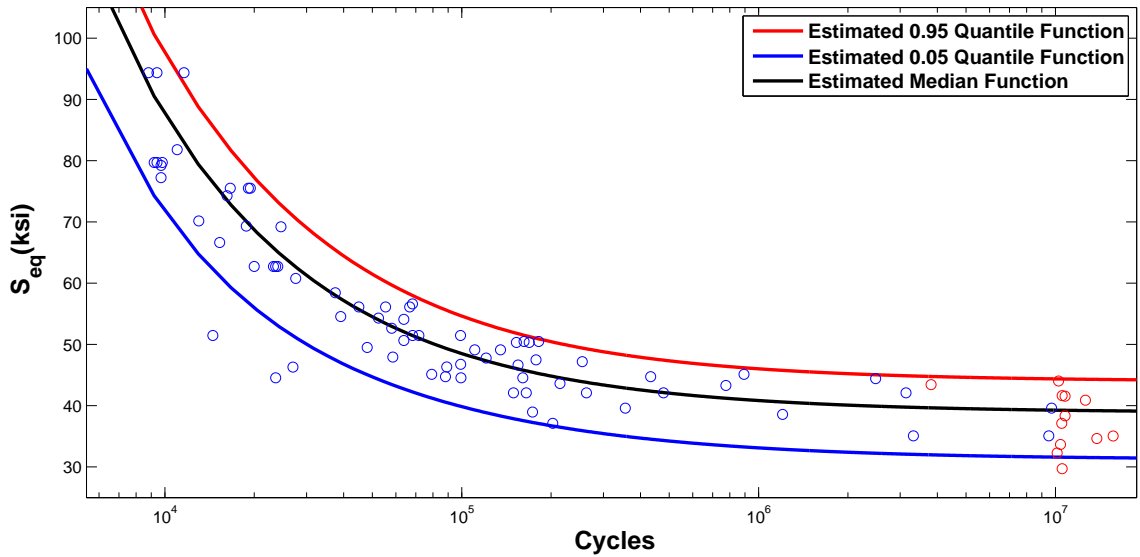}
\caption{Model IId fit of the 75S-T6 data set.  Allowing non-constant variance in a random fatigue-limit model for censored data has the effect of slightly reducing the gap between the median and the $0.95$ and $0.05$ quantiles for the highest values of $S_{eq}$.}
\label{fit2d}
\end{figure} 

Table \ref{mle2} shows the maximum likelihood estimates and the maximum likelihood values obtained for models IIc and  IId. Again, the random fatigue-limit model with the sev distribution (Model IId) performs slightly better than the random fatigue-limit model with the lognormal distribution (Model IIc). The corresponding fit of Model IId in Figure \ref{fit2d} is very similar to the one obtained by Model IIb in Figure \ref{fit2b}. 

\begin{table}[h!]
\begin{center}
\caption{Maximum likelihood estimates for Model IIc and Model IId.}
\begin{tabular}{|c|c|c|c|c|c|c|c|c|}
\hline
 & $A_1$ & $A_2$ & $\mu_f$ & $\sigma_f$ & $q$ & $B_1$ & $B_2$ & $\log(L^*)$\\
\hline
Model IIc & 6.43 & -1.44 & 1.58 & 0.0408 & 0.4923 & 2.68 & -1.97 & -908.15 \\
\hline
Model IId & 6.49 & -1.46 & 1.60 & 0.0366 & 0.4904 & 0.66 & -0.94 & -906.73\\
\hline
\end{tabular}
\label{mle2}
\end{center}
\end{table}

\begin{figure}[h!]
\centering
\begin{minipage}{.45\textwidth}
  \centering
  \includegraphics[width=9cm]{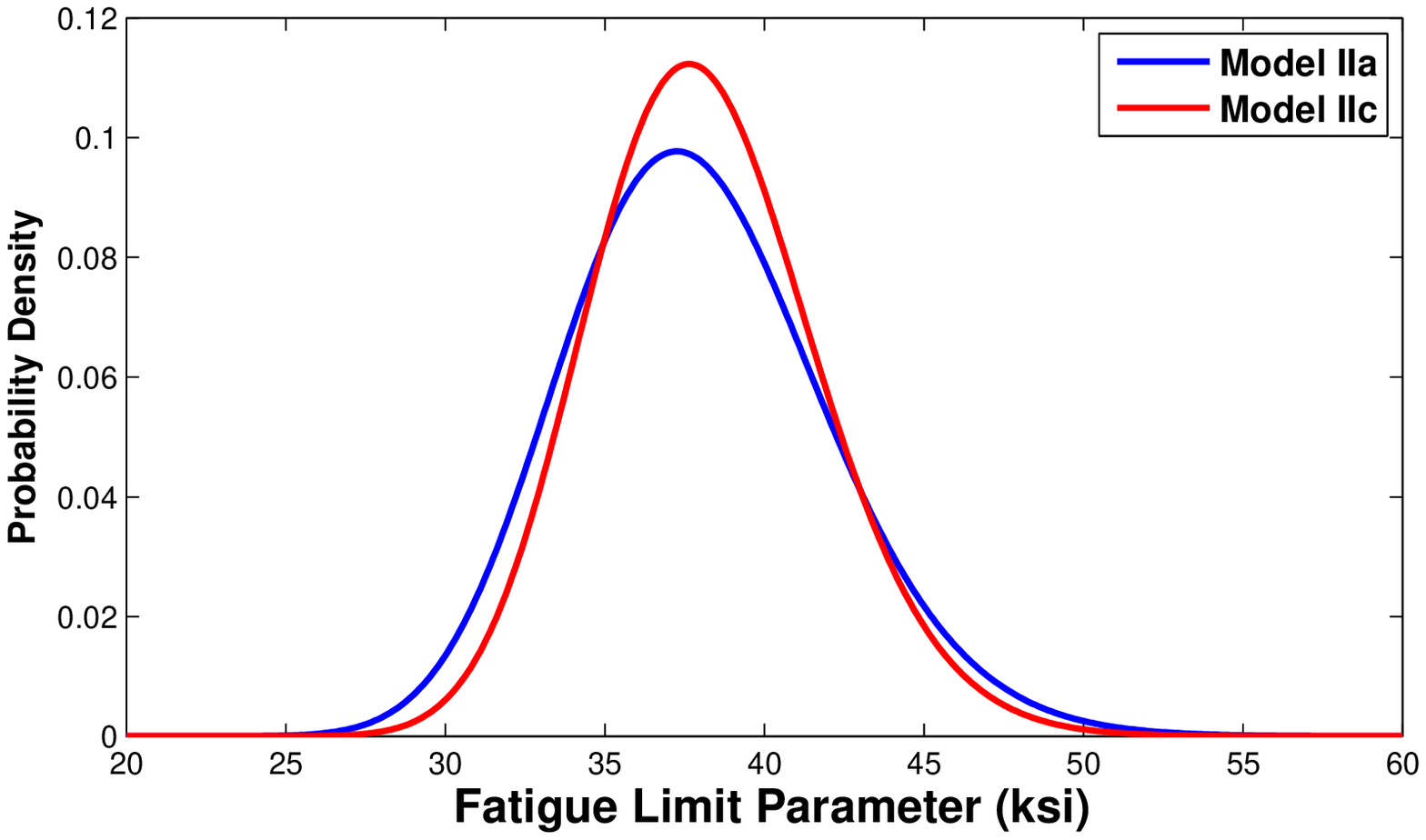}
  \captionof{figure}{Estimated probability density functions of the fatigue limit parameter, $A_3$, for models IIa and IIc.}
  \label{FLpdf1}
\end{minipage}
~
\begin{minipage}{.45\textwidth}
  \centering
  \includegraphics[width=9cm]{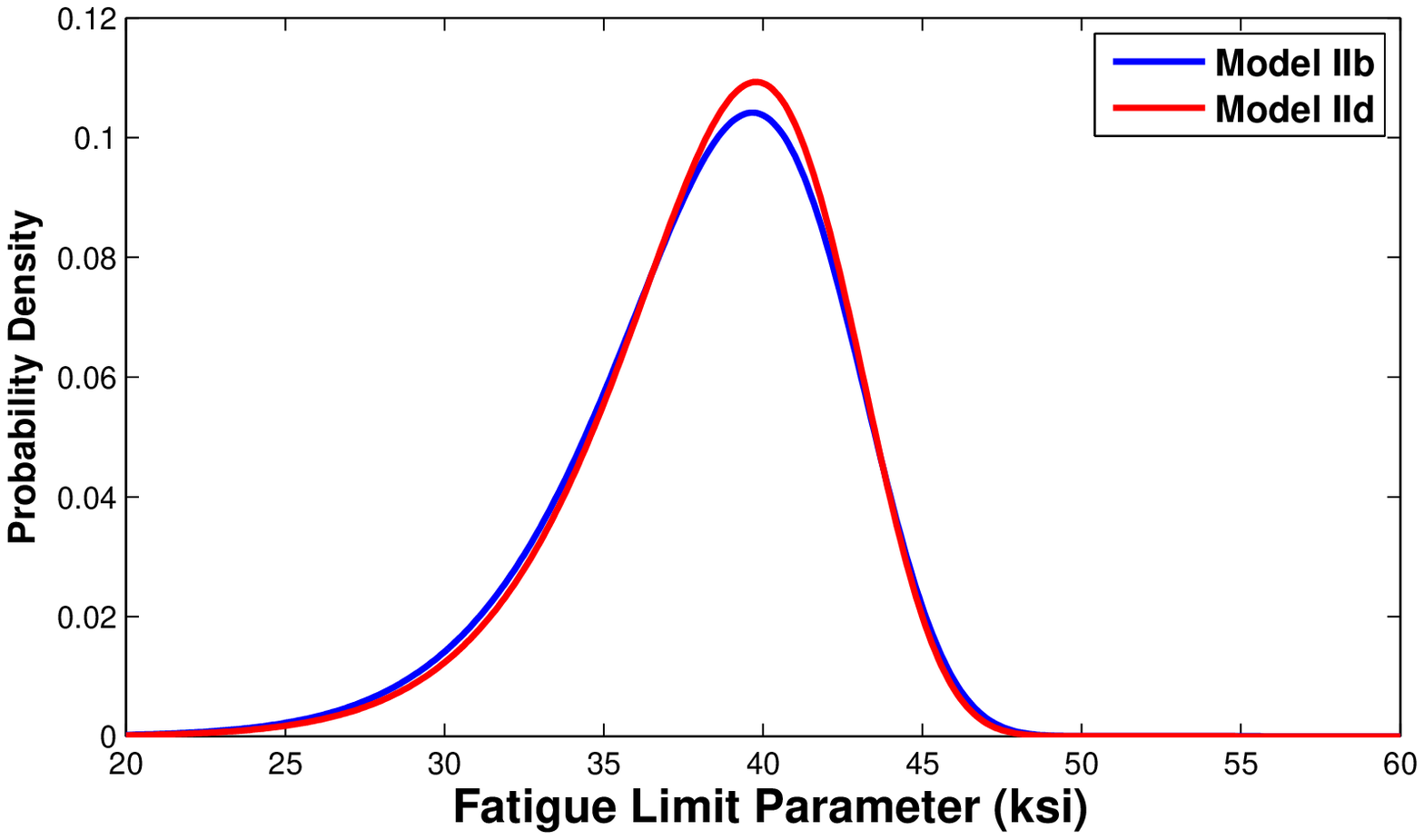}
  \captionof{figure}{Estimated probability density functions of the fatigue limit parameter, $A_3$, for models IIb and IId..}
  \label{FLpdf2}
\end{minipage}
\end{figure}

Figures \ref{FLpdf1} and \ref{FLpdf2} show the probability density function for $A_3$ corresponding to models IIa, IIb, IIc and IId.

In the next subsections, our goal is to compare the relative performances of the proposed models that include an adequate formulation in terms of run-outs. As an initial step, we explore the consistency of the fitted models by looking at the variability in the confidence bands of the quantile functions of fatigue life.

\subsection{Bootstrap confidence bands and confidence intervals} 
\label{sec3b}
We obtain bootstrap confidence bands for the model fittings Ia, Ib, IIb and IId, as illustrated in Figures \ref{fit1}, \ref{fit1b}, \ref{fit2b} and \ref{fit2d}, respectively. Stratified bootstrap algorithm \ref{alg2} is implemented with censored data. First, the data set is stratified on the basis of the cycle ratio, $R$. Then, we sample independently from each stratum where each sample contains $S_{max}, R, N$ and the binary variable $\delta$ (See \cite{efron}). By repetition, we generate $M = 200$ bootstrap data sets. For each data set, we obtain the maximum likelihood estimate and compute the corresponding quantiles. 

\begin{algorithm}[h!]
\caption{Stratified bootstrap algorithm for censored data} \label{alg2}
\begin{algorithmic}[1]
\State \textbf{set} $ data = [data_1, data_2, \ldots, data_n] $
\For{$ i = 1$ : $n$} 
\State \textbf{draw} $|data_i|$ samples with replacement from $data_i$
\State \textbf{let} $data_i^*$ be the bootstrap stratum. 
\EndFor
\State \textbf{let} $data^{*} = [data^*_1, data^*_2, \ldots, data^*_n]$ be the bootstrap data set.
\State \textbf{find} the maximum likelihood estimate $\theta^*$ given $data^{*}$
\State \textbf{compute} the bootstrap quantiles
\State \textbf{repeat} steps (2 to 7) $M$ times.
\end{algorithmic}
\end{algorithm}

\begin{figure}[h!]
\centering
\includegraphics[width=18cm]{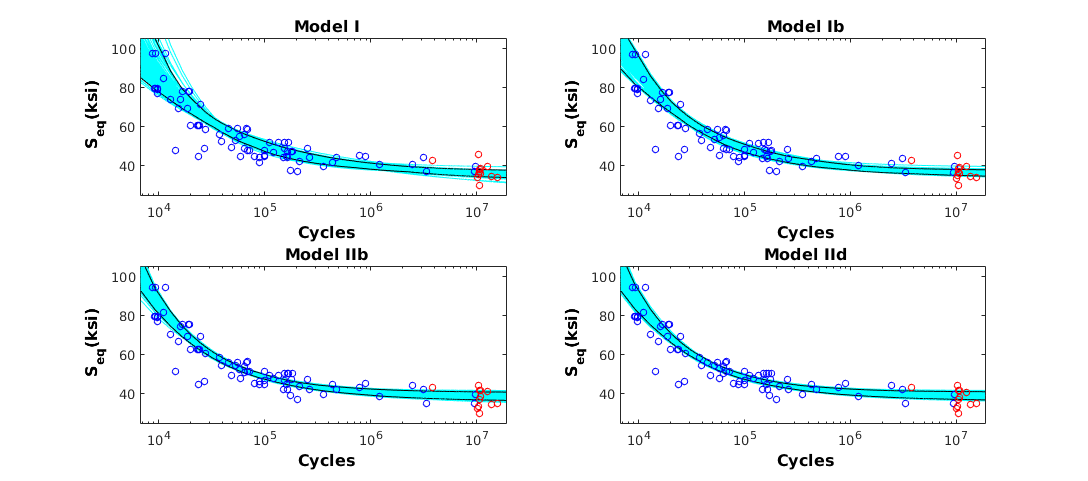}
\caption{$95\%$ bootstrap confidence bands for the median of fatigue life.}
\label{boots1}
\end{figure}

\begin{figure}[h!]
\centering
\includegraphics[width=18cm]{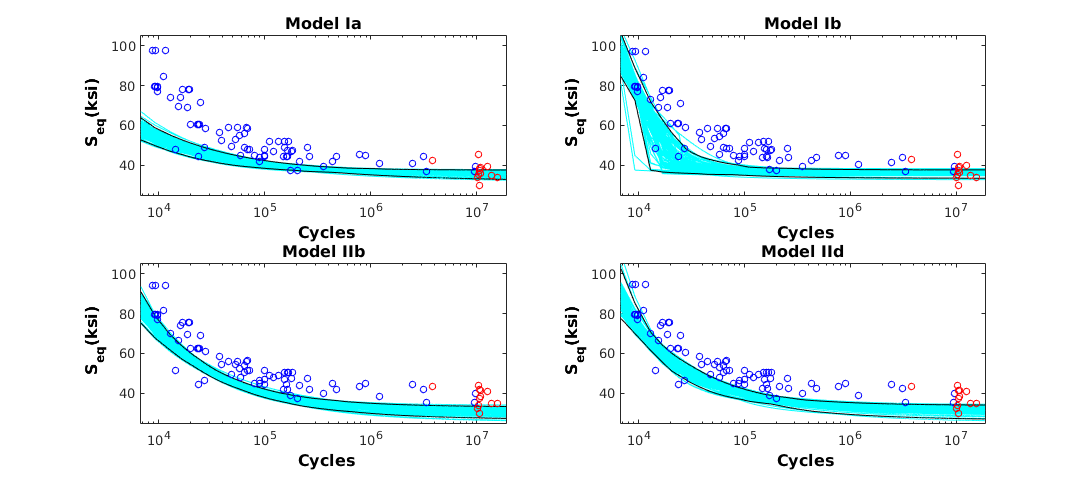}
\caption{$95\%$ bootstrap confidence bands for the 0.05 quantile of fatigue life. The 0.05 quantile is not as robust as the median, especially for Model Ib.}
\label{boots2}
\end{figure}   

Figure \ref{boots1} shows the median functions (blue curves) and the bootstrapped $95$\% confidence bands (black curves) for models Ia, Ib, IIb and IId. Figure \ref{boots2} shows the $0.05$ quantiles (blue curves) and the bootstrapped $95$\% confidence bands (black curves). Table \ref{boots3} provides the bootstrap confidence intervals for the maximum likelihood estimates for these models. Clearly, the random fatigue-limit models (Model IIb and Model IId) provide the narrowest confidence intervals for $A_1, A_2$ and $q$. 

\begin{table}[h!]
\begin{center}
\caption{$95\%$ bootstrap confidence intervals for the maximum likelihood estimates.}
\begin{tabular}{|c|c|c|c|c|c|c|c|}
\hline
\multicolumn{7}{|c|}{Model Ia} \\
\hline
$A_1$ & $A_2$ & $A_3$ & $q$ & \multicolumn{3}{|c|}{$\tau$} \\
\hline
(6.19, 8.79) &  (-2.88, -1.22) & (31.01, 38.46) & (0.487, 0.613) &   \multicolumn{3}{|c|}{ (0.355, 0.646)}  \\
\hline
\multicolumn{7}{|c|}{Model Ib} \\
\hline
$A_1$ & $A_2$ & $A_3$ & $q$ & $B_1$ & \multicolumn{2}{|c|}{$B_2$} \\
\hline
(6.28, 7.45) &  (-2.05, -1.31) & (33.66, 38.33) & (0.460, 0.595) & (3.48, 6.25) &  \multicolumn{2}{|c|}{(-3.92, -2.31)}  \\
\hline
\multicolumn{7}{|c|}{Model IIb} \\
\hline
$A_1$ & $A_2$ & $\mu_f$ & $\sigma_f$ & $q$ & \multicolumn{2}{|c|}{$\tau$} \\
\hline
(6.23, 6.87) &  (-1.70, -1.30) & (1.58, 1.62) & (0.0275, 0.0497) & (0.451, 0.515) &  \multicolumn{2}{|c|}{(0.035, 0.123)}  \\
\hline
\multicolumn{7}{|c|}{Model IId} \\
\hline
$A_1$ & $A_2$ & $\mu_f$ & $\sigma_f$ & $q$ & $B_1$ & $B_2$ \\
\hline
(6.21, 6.89) &  (-1.71, -1.29) & (1.58, 1.62) & (0.0240, 0.0476) & (0.456, 0.519) & (-7.43, 4.43) & (-3.11, 3.36)  \\
\hline
\end{tabular}
\label{boots3}
\end{center}
\end{table}

\subsection{Model comparison}
\label{sec3c}
Using a classical approach, we compute some popular information criteria, such as Akaike information criterion (AIC) \cite{aic}, Bayesian information criterion (BIC) \cite{bic1,bic2} and AIC with correction \cite{aicc}, which are based on the maximized log-likelihood values. Such measures take into account both the goodness of fit and the complexity of the models in terms of the number of parameters.

Table \ref{Capp} contains the maximum log-likelihood values that correspond to the models introduced in Subsections \ref{model1} -- \ref{model2d} together with the classical information criteria computations. These classical evaluations of model uncertainty indicate that, despite its complexity,  Model IIb is preferable.

\begin{table}[h!]
\begin{center}
\caption{Classical information criteria show that Model IIb provides the best fit to the 75S-T6 data set.}
\begin{tabular}{|c|c|c|c|c|c|c|c|}
\hline
Models &  \bf{Ia} & \bf{Ib} & \bf{IIa} & \bf{IIb} & \bf{IIc} & \bf{IId} \\
\hline
maximum log-likelihood & -950.16 & -920.51 &  -913.42  &  -907.31 & -908.15 & -906.73 \\
\hline
Akaike Information Criterion (AIC) & 1910.3 & 1853.0 &  1838.8  &  1826.6 &  1830.3 & 1827.5   \\
\hline
Bayesian Information Criterion (BIC) & 1922.5 & 1867.7 &  1853.5  &  1841.3   & 1847.4 & 1844.6      \\
\hline
Akaike Information Criterion with correction & 1911.1 & 1854.1  & 1839.9 &  1827.7  &  1831.8 & 1828.9  \\
\hline
\end{tabular}
\label{Capp}
\end{center}
\end{table}

\section{Bayesian approach}
\label{sec4}
\subsection{Model calibration}
\label{sec4a}
We consider now a Bayesian approach to study models Ia, Ib and IIb under two different scenarios. For each scenario, we compute the maximum posterior estimate (analytically) using the Laplace method and provide Markov chain Monte Carlo (MCMC) posterior samples. The random walk Metropolis-Hastings algorithm (\ref{alg1}) is used to generate MCMC samples. We use a normal proposal distribution to perturb the current simulated vector, $\theta_c$, and generate a new perturbed vector, $\theta_p \sim N(\theta_c, diag(\delta))$, where $\delta$ is a vector of parameters that controls the acceptance rate of the algorithm. After several attempts, we chose $\delta$ such that we could obtain a reasonable acceptance rate (see \cite[Chapter 6]{robert}). 

\begin{algorithm}[h!]
\caption{Random walk Metropolis-Hastings algorithm} \label{alg1}
\begin{algorithmic}[1]
\State \textbf{set} an initial value for the chain: $\theta_c = \theta_0$ and \textbf{choose} $\delta$
\State \textbf{compute} $a = loglikelihood(\theta_c) + logprior(\theta_c)$
\State \textbf{draw} $\theta_p$ from $N(\theta_c, diag(\delta))$
\State \textbf{compute} $b = loglikelihood(\theta_p) + logprior(\theta_p)$
\State \textbf{let} $ H = min(1,exp(b-a))$ and \textbf{draw} $r$ from $U(0,1)$
\If{$H > r$}
\State $\theta_c = \theta_p$
\State $ a = b $
\EndIf
\State \textbf{repeat} steps (3 to 8) \textbf{until} $L$ posterior samples are accepted.
\end{algorithmic}
\end{algorithm}

For both scenarios, the algorithm is initialized as follows:

\begin{itemize}
\item Model Ia: $\theta_0 = (7.4, -2, 35, 0.56, 0.5)$ and $\delta = (0.1, 0.1, 0.1, 0.01, 0.05)$. 
\item Model Ib: $\theta_0 = (6.7, -1.6, 36.2,  0.55,  4.6,  -2.9)$ and $\delta = (0.1, 0.1, 0.1, 0.01, 0.1, 0.1)$. 
\item Model IIb: $\theta_0 = (6.5, -1.5, 1.6,  0.04, 0.49, 0.085)$ and $\delta = (0.1, 0.1, 0.005, 0.001, 0.01, 0.01)$.
\end{itemize}

Each chain was run for $1,010,000$ times, with a 10,000 iterations burn-in period and every 50th draw of the chain kept. The MCMC posterior samples were summarized by the Laplace-Metropolis estimator (see \cite{lewraf}), the empirical mean and standard deviation and the estimated marginal densities. The marginal densities were obtained by kernel density estimation (KDE) with a normal kernel function. The bandwidth was chosen to be optimal for normal densities. 

In attempting to provide an objective Bayesian analysis, we considered two different scenarios. For both scenarios, we chose data-dependent proper priors \cite{berger}. In the first scenario, normal priors centered around the maximum likelihood estimates with arbitrary variance were considered for all the parameters except the standard deviations that were assigned inverse-gamma priors. The second scenario adopted less informative uniform priors for all the parameters. The uniform priors were chosen by spreading the range of the likelihood function then tuning these priors until we obtained proper untruncated posterior distributions.

\subsubsection{Scenario 1 (informative priors)}

In scenario 1, we considered the following informative priors that were induced from the maximum likelihood estimates as explained previously.

\begin{itemize}
\item Model Ia:  $A_1 \sim \mathcal{N}(7.4, 2)$, $A_2 \sim \mathcal{N}(-2, 2)$, $A_3 \sim \mathcal{N}(35, 2)$, $q \sim \mathcal{N}(0.56, 0.5)$, $\tau \sim \mathcal{IG}(0.5,0.25)$.  
\item Model Ib:  $A_1 \sim \mathcal{N}(6.7, 2)$, $A_2 \sim \mathcal{N}(-1.6, 2)$, $A_3 \sim \mathcal{N}(36.2, 2)$, $q \sim \mathcal{N}(0.55, 0.5)$, $B_1 \sim \mathcal{N}(4.6, 2)$, $B_2 \sim \mathcal{N}(-2.9, 2)$.
\item Model IIb:  $A_1 \sim \mathcal{N}(6.5, 2)$, $A_2 \sim \mathcal{N}(-1.5, 2)$, $\mu_{f} \sim \mathcal{N}(1.6, 0.1)$,  $\sigma_{f} \sim \mathcal{IG}(2, 0.1)$, $q \sim \mathcal{N}(0.49, 0.5)$,  $\tau \sim \mathcal{IG}(1, 0.1)$.  
\end{itemize}

\textbf{Numerical Results - Model Ia }

\begin{figure}[h!]
\centering
\includegraphics[width=17cm]{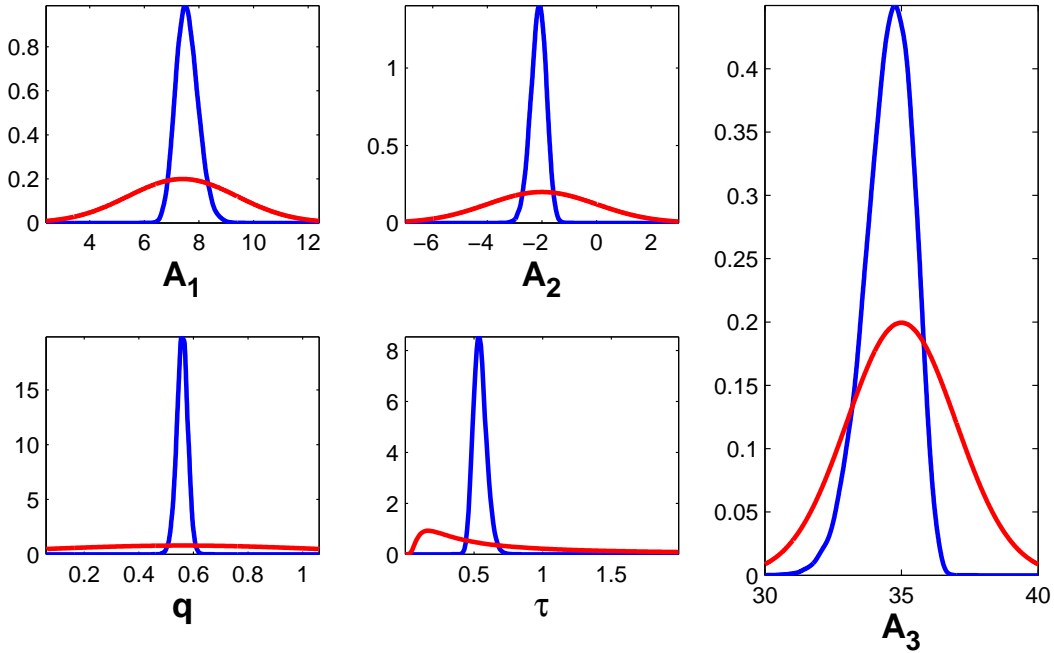}
\caption{Prior densities (red line) and approximate marginal posterior densities (blue line) for $A_1, A_2, q, \tau$ and $A_3$. The marginal posterior densities for all parameters are highly concentrated around their unique mode, suggesting that the observed data, given the assumed model, considerably increase our degree of belief about the range of the parameters. The high concentrations of $q$ and $\tau$ are especially noticeable. The estimated marginal posterior of the fatigue limit parameter, $A_3$, is left-skewed although the prior was assumed to be a normal distribution.}
\label{MCMC1}
\end{figure}   

\begin{table}[h!]
\begin{center}
\caption{Maximum posterior estimates for Model Ia.}
\begin{tabular}{|c|c|c|c|c|c|}
\hline
Estimator & $A_1$ & $A_2$ & $A_3$ & $q$ & $\tau$ \\
\hline
Laplace & 7.39 & -2.01 & 35.03 & 0.563 & 0.523 \\
\hline
Laplace-Metropolis & 7.46 & -2.07 & 34.92 & 0.561 & 0.524 \\
\hline
\end{tabular}
\label{MAP1}
\end{center}
\end{table}
 
\begin{table}[h!]
\begin{center}
\caption{MCMC posterior empirical mean estimates with their standard deviations.} 
\begin{tabular}{|c|c|c|c|c|c|}
\hline
 \, & $A_1$ & $A_2$ & $A_3$ & $q$ & $\tau$ \\
\hline   
 Mean & 7.57 & -2.13 & 34.53 & 0.559 & 0.544 \\
\hline    
 SD & 0.41 & 0.28 & 0.88 & 0.020 & 0.048 \\
\hline
\end{tabular}
\label{mean1}
\end{center}
\end{table} 

\begin{figure}[h!]
\centering
\includegraphics[width=22cm]{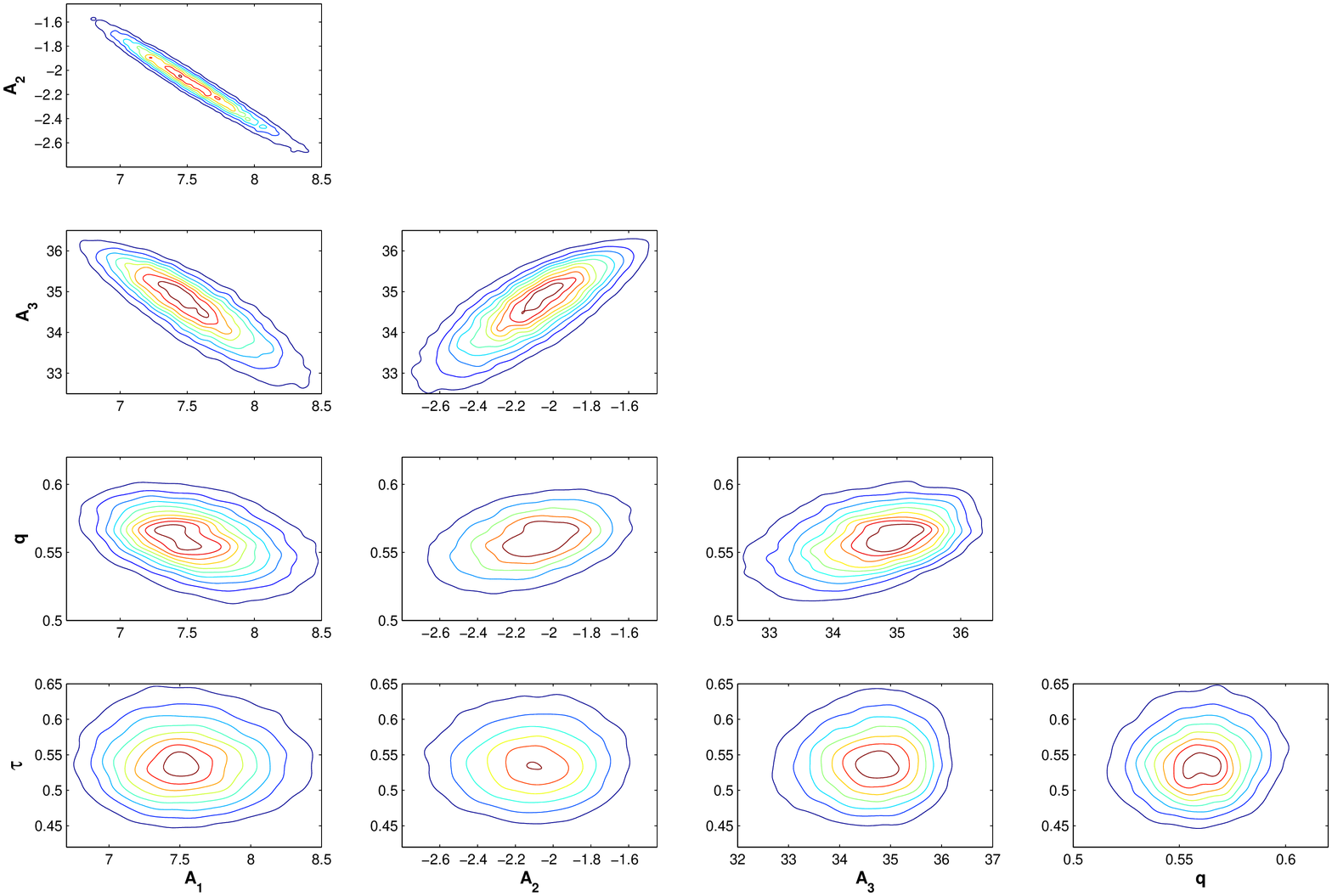}
\caption{Contour plots of the estimated bivariate densities for each pair of parameters in Model Ia. A strong correlation appears between $A_1$ and $A_2$ and they also appear to be linearly dependent. The fatigue limit parameter, $A_3$, is highly correlated with $A_1$ and $A_2$.}
\label{biv1}
\end{figure}

Maximum posterior estimates shown in Table \ref{MAP1} are similar to the maximum likelihood estimates obtained for Model Ia (Table \ref{mle1a}). Figure \ref{MCMC1} and empirical standard deviations given in Table \ref{mean1} show that the fatigue limit parameter, $A_3$, is the most uncertain parameter whereas $q$ is the least uncertain parameter. Figure \ref{MCMC1} also shows that the data are informative for all the parameters because there is a contraction between the prior densities and the posterior densities. Correlation coefficients presented in Table \ref{Tbiv1} and Figure \ref{biv1} show that $A_1$ and $A_2$ are approximately linear dependent. We can therefore reduce the number of parameters in Model Ia by one parameter. These parameters are also highly correlated with the fatigue limit parameter, $A_3$. On the other hand, there is a weak linear relationship between $q$ and the parameters $A_1, A_2$ and $A_3$. Moreover, the standard deviation, $\tau$, has no notable correlation with any parameter.

\begin{table}[h!]
\begin{center}
\caption{Correlation coefficients for each pair of parameters in Model Ia.} 
\begin{tabular}{|c|c|c|c|c|}
\hline
   & $A_1$ & $A_2$ & $A_3$ & $q$ \\
\hline   
$A_2$ & -0.980 & --- & --- & ---  \\
\hline    
$A_3$ & -0.860 & 0.799 &  ---  & ---  \\
\hline
 $q$ & -0.385 & 0.365 & 0.384 & ---  \\
\hline
$\tau$ & -0.005 & 0.003 & 0.050 & 0.073 \\
\hline
\end{tabular}
\label{Tbiv1}
\end{center}
\end{table} 

\newpage \textbf{Numerical Results - Model Ib }

\begin{figure}[h!]
\centering
\includegraphics[width=17cm]{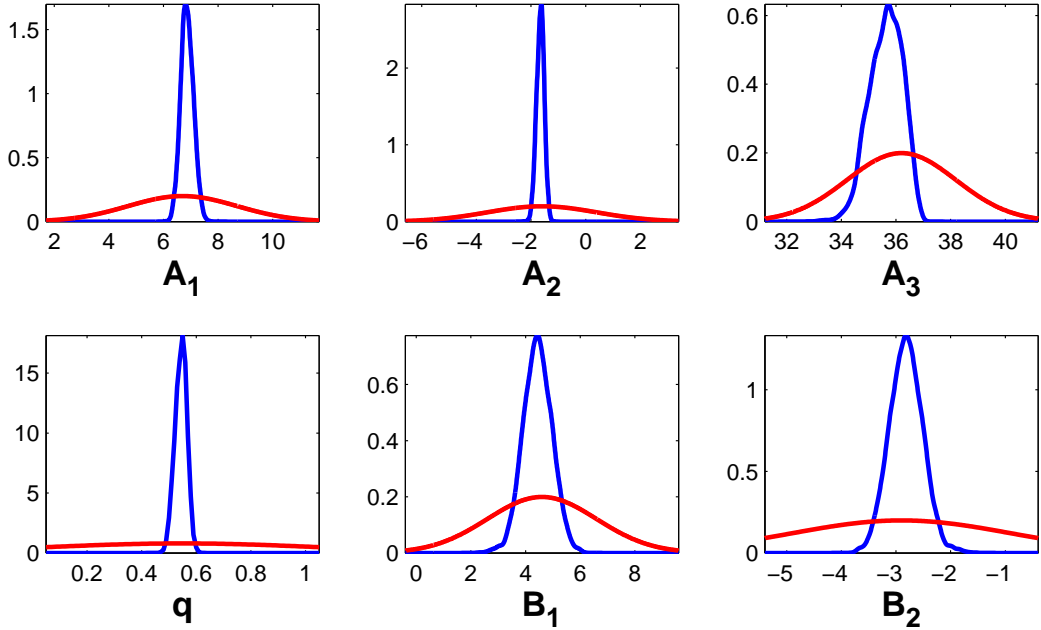}
\caption{Prior densities (red line) and approximate marginal posterior densities (blue line) for $A_1, A_2, q, B_1, B_2$ and $A_3\,.$ The estimated posterior densities for all parameters are more concentrated than the prior densities, which means that the data are informative. Again, the estimated posterior of $q$ is highly concentrated. Allowing a non-constant variance has the effect of reducing the uncertainties of $A_1, A_2$ and the fatigue limit parameter, $A_3$. The estimated marginal posterior of the fatigue limit parameter is left-skewed although the prior was assumed to be a normal distribution.}
\label{MCMC1b}
\end{figure}  

\begin{table}[h!]
\begin{center}
\caption{Maximum posterior estimates for Model Ib.}
\begin{tabular}{|c|c|c|c|c|c|c|}
\hline
Estimator & $A_1$ & $A_2$ & $A_3$ & $q$ & $B_1$ & $B_2$ \\
\hline
Laplace & 6.72 & -1.57 & 36.21 & 0.551 & 4.56 & -2.89 \\
\hline
Laplace-Metropolis & 6.78 & -1.61 & 36.20 & 0.552 & 4.43 & -2.83 \\
\hline
\end{tabular}
\label{MAP1b}
\end{center}
\end{table}
 
\begin{table}[h!]
\begin{center}
\caption{MCMC posterior empirical mean estimates with their standard deviations.} 
\begin{tabular}{|c|c|c|c|c|c|c|}
\hline
 \, & $A_1$ & $A_2$ & $A_3$ & $q$ & $B_1$ & $B_2$ \\
\hline   
 Mean & 6.87 & -1.66 & 35.63 & 0.544 & 4.44 & -2.81 \\
\hline    
 SD & 0.23 & 0.14 & 0.60 & 0.022 & 0.53 & 0.31 \\
\hline
\end{tabular}
\label{mean1b}
\end{center}
\end{table} 

\begin{figure}[h!]
\centering
\includegraphics[width=22cm]{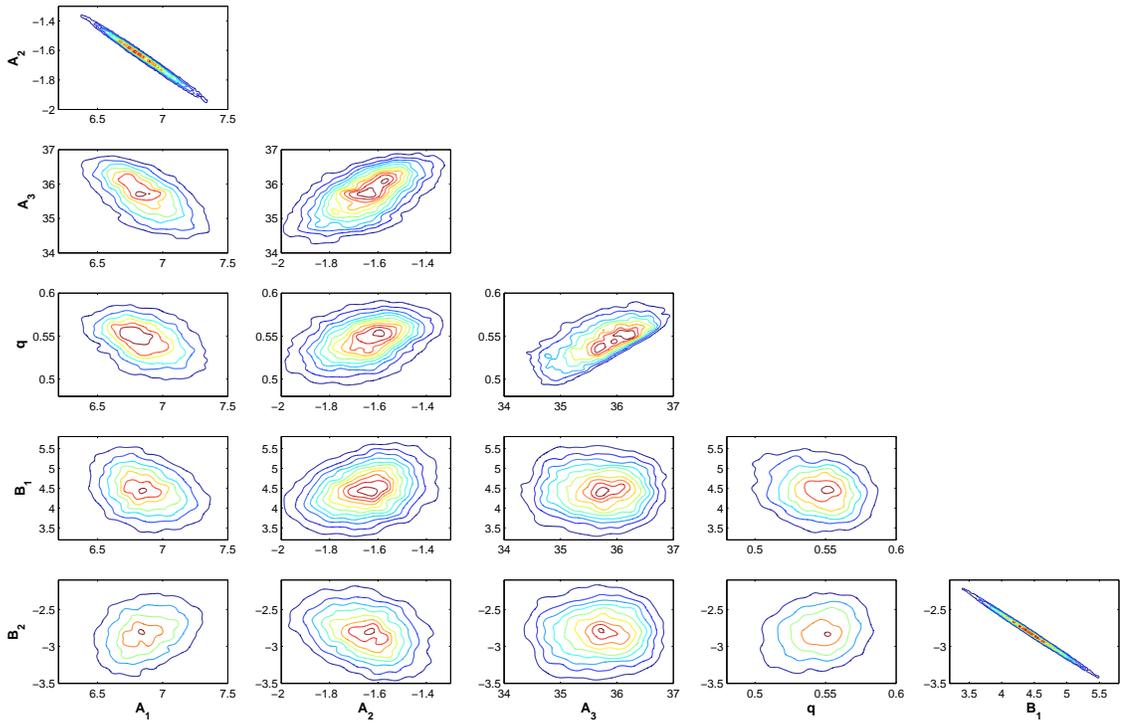}
\caption{Contour plots of the approximate bivariate densities for each pair of parameters in Model Ib. There are two strong correlations between $A_1$ and $A_2$ and between $B_1$ and $B_2$. Such strong correlation suggests linear dependence; it is therefore possible to remove two parameters from Model Ib. The fatigue limit parameter, $A_3$, shows a moderate correlation with $A_1, A_2$ and $q$. Allowing a non-constant variance has the effect of increasing the correlation between $q$ and the fatigue limit parameter, $A_3$.}
\label{biv1b}
\end{figure}

Maximum posterior estimates given in Table \ref{MAP1b} are similar to the maximum likelihood estimates obtained for Model Ib (Table \ref{mle1b}). Similarly to Model Ia, Figure \ref{MCMC1b} and Table \ref{mean1b} show that the fatigue limit parameter, $A_3$, is the most uncertain parameter whereas $q$ is the least uncertain parameter. However, the uncertainties have been reduced for $A_1, A_2$ and $A_3$ when compared with Model Ia. Figure \ref{MCMC1b} shows again that the data are informative for all the parameters as previously explained. The marginal posterior of the fatigue limit parameter, $A_3$ is left-skewed similar to the profile likelihood estimate. Correlation coefficients shown in Table \ref{Tbiv1b} and Figure \ref{biv1b} show that $A_1$ and $B_1$ are almost perfectly correlated with $A_2$ and $B_2$, respectively. Thus, we can consider a fatigue limit model with non-constant variance with only four parameters, which is the same number of parameters in the logarithmic fit. The fatigue limit parameter in Model Ib has a moderate linear relationship with $A_1, A_2$ and $q$ whereas the fatigue limit parameter in Model Ia has a strong linear relationship with $A_1$ and $A_2$ and a weak linear relationship with $q$. 

\begin{table}[h!]
\begin{center}
\caption{Correlation coefficients for each pair of parameters in Model Ib.} 
\begin{tabular}{|c|c|c|c|c|c|}
\hline
   & $A_1$ & $A_2$ & $A_3$ & $q$ & $B_1$ \\
\hline   
$A_2$ & -0.993 & --- & --- & --- & --- \\
\hline    
$A_3$ & -0.610 &0.592 & --- & --- & --- \\
\hline
 $q$ & -0.384 & 0.396 & 0.658 & --- & --- \\
\hline
$B_1$ & -0.301 & 0.308 & 0.017 & -0.177 & ---\\
\hline
$B_2$ & 0.300 & -0.306 & -0.011 & 0.188 & -0.997 \\
\hline
\end{tabular}
\label{Tbiv1b}
\end{center}
\end{table} 

\newpage \textbf{Numerical Results - Model IIb}

\begin{figure}[h!]
\centering
\includegraphics[width=17cm]{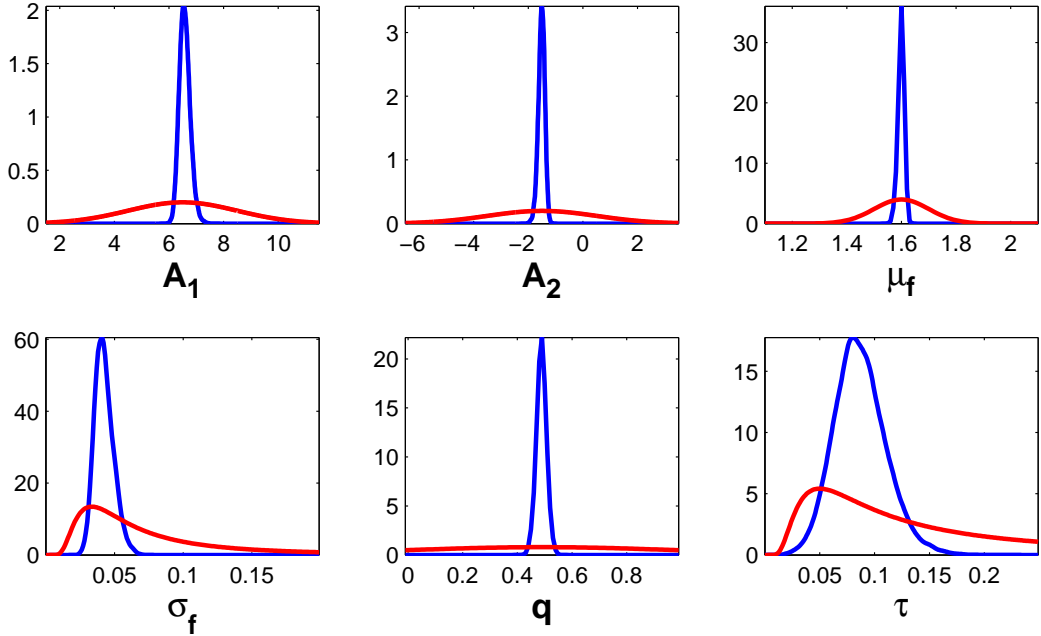}
\caption{Prior densities (red line) and approximate marginal posterior densities (blue line) for $A_1, A_2, q,\tau, \mu_{f}$ and $\sigma_{f}$. The posterior densities for all parameters are more concentrated than the prior densities, which means the data are informative. The high concentrations of the location and scale parameters, $\mu_f$ and $\sigma_f$, are particularly noticeable. The random fatigue-limit model has the effect of considerably reducing the uncertainties of $A_1, A_2$ and $\tau$.}
\label{MCMCFL}
\end{figure}   

\begin{table}[h!]
\begin{center}
\caption{Maximum posterior estimates for Model IIb.}
\begin{tabular}{|c|c|c|c|c|c|c|}
\hline
Estimator & $A_1$ & $A_2$ & $\mu_{f}$ & $\sigma_{f}$ & $q$ & $\tau$ \\
\hline
Laplace & 6.51 & -1.47 & 1.60 & 0.0387 & 0.488 & 0.082 \\
\hline
Laplace-Metropolis & 6.53 & -1.49 & 1.60 & 0.0386 & 0.485 & 0.080 \\
\hline
\end{tabular}
\label{MAPFL}
\end{center}
\end{table}
 
\begin{table}[h!]
\begin{center}
\caption{MCMC posterior empirical mean estimates with their standard deviations.} 
\begin{tabular}{|c|c|c|c|c|c|c|}
\hline
 \, & $A_1$ & $A_2$ & $\mu_{f}$ & $\sigma_{f}$ & $q$ & $\tau$ \\
\hline   
 Mean & 6.58 & -1.52 & 1.60 & 0.0424 & 0.488 & 0.087\\
\hline    
 SD & 0.20 & 0.12 & 0.012 & 0.007  &  0.018 & 0.023 \\
\hline
\end{tabular}
\label{meanFL}
\end{center}
\end{table} 

\begin{figure}[h!]
\centering
\includegraphics[width=22cm]{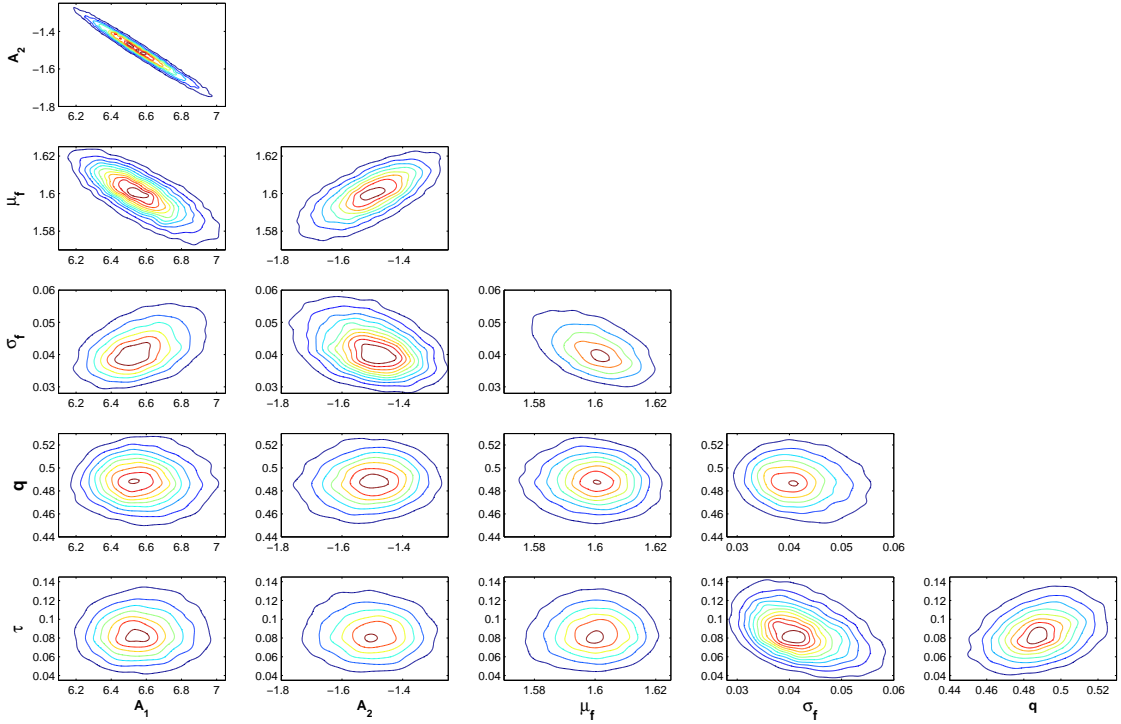}
\caption{Contour plots of the approximate bivariate densities for each pair of parameters in Model IIb. Again, a strong correlation appears between $A_1$ and $A_2$. Also, the parameter $\mu_f$ has a relatively strong correlation with $A_1$ and $A_2$. The random fatigue-limit model has the effect of reducing the correlations between $q$ and the parameters $A_1$ and $A_2$.}
\label{bivFL}
\end{figure}

Maximum posterior estimates presented in Table \ref{MAPFL} are similar to the maximum likelihood estimates obtained for Model IIb (Table \ref{mle1}). Figure \ref{MCMCFL} and Table \ref{meanFL} show that the location and scale parameters, $\mu_f$ and $\sigma_f$, are the least uncertain parameters. Moreover, the uncertainties have been reduced for $A_1, A_2$ and $q$ when compared with Model Ia and Model Ib. Figure \ref{MCMCFL} shows that the data are very informative for all the parameters because there is a strong contraction between the prior densities and the posterior densities. Similarly to Model Ia, Table \ref{TbivFL} and Figure \ref{bivFL} show that $A_1$ and $A_2$ are approximately linear dependent, and therefore we can reduce the number of parameters for Model IIb by one parameter. The location parameter, $\mu_f$, is strongly correlated with $A_1$ and $A_2$ whereas $\sigma_f$ is moderately correlated with $A_1, A_2$ and $\mu_f$. There is a weak negative correlation between $\tau$ and $\sigma_f$ and a weak positive correlation between $\tau$ and $q$. 

\begin{table}[h!]
\begin{center}
\caption{Correlation coefficients for each pair of parameters in Model IIb.} 
\begin{tabular}{|c|c|c|c|c|c|}
\hline
  & $A_1$ & $A_2$ & $\mu_{f}$ & $\sigma_{f}$ & $q$ \\
\hline   
$A_2$ & -0.986 & --- & --- & --- & --- \\
\hline    
 $\mu_{f}$  & -0.777 & 0.708 &  ---  & ---  & --- \\
\hline
 $\sigma_{f}$ & 0.447 & -0.404 & -0.526 & --- & --- \\
\hline
$q$ &  -0.045 & 0.090 & -0.062 & -0.145 & --- \\
\hline
$\tau$ &  0.034  & -0.022 & 0.042 & -0.396 & 0.321 \\
\hline
\end{tabular}
\label{TbivFL}
\end{center}
\end{table} 

\subsubsection{Scenario 2 (uninformative priors)}

Now, we provide the same results but for prescribed uninformative priors as follows: 
\begin{itemize}
\item Model Ia:  $A_1 \sim U(5, 13)$, $A_2 \sim U(-5, 0)$, $A_3 \sim U(24, 40)$, $q \sim U(0.1, 0.9)$, $\tau \sim U(0.1, 1.5)$.
\item Model Ib:  $A_1 \sim U(4, 10)$, $A_2 \sim U(-4, 0)$, $A_3 \sim U(30, 40)$, $q \sim U(0.1, 0.9)$, $B_1 \sim U(2, 7)$, $B_2 \sim U(-5, -1)$.
\item Model IIb:  $A_1 \sim U(4, 10)$, $A_2 \sim U(-4, 0)$, $\mu_{f} \sim U(1.4, 1.8)$, $\sigma_{f} \sim U(0,0.1)$, $q \sim U(0.1, 0.9)$, $\tau \sim U(0, 0.25)$.  
\end{itemize}

\textbf{Numerical Results - Model Ia}

\begin{figure}[h!]
\centering
\includegraphics[width=17cm]{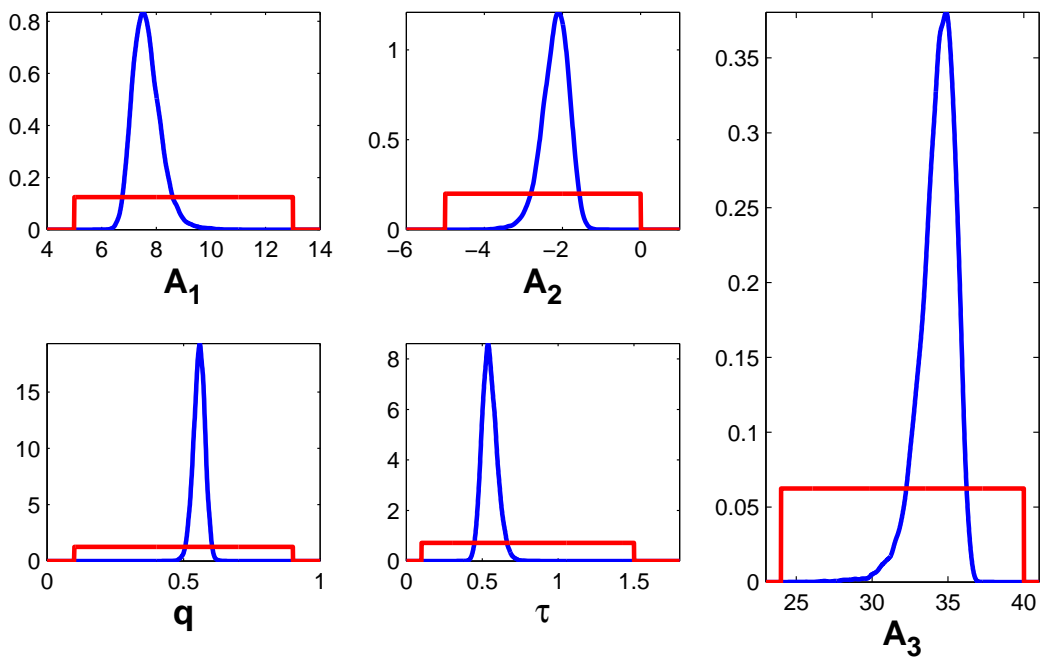}
\caption{Prior densities (red line) and approximate marginal posterior densities (blue line) for $A_1, A_2, q, \tau$ and $A_3\,.$ The estimated posterior densities for all parameters are concentrated around the corresponding modes. Again, we notice the high concentrations of $q$ and $\tau$ as in Scenario 1. The estimated marginal posteriors of $A_1, A_2$ and $A_3$ show greater variability in comparison with Figure \ref{MCMC1}.}
\label{uMCMC1}
\end{figure}   

\begin{table}[h!]
\begin{center}
\caption{Maximum posterior estimates for Model Ia.}
\begin{tabular}{|c|c|c|c|c|c|}
\hline
Estimator & $A_1$ & $A_2$ & $A_3$ & $q$ & $\tau$ \\
\hline
Laplace & 7.38 & -2.01 & 35.04 & 0.563 & 0.527 \\
\hline
Laplace-Metropolis & 7.39 & -2.02 & 35.07 & 0.563 & 0.517 \\
\hline
\end{tabular}
\label{uMAP1}
\end{center}
\end{table}
 
\begin{table}[h!]
\begin{center}
\caption{MCMC posterior empirical mean estimates with their standard deviations.} 
\begin{tabular}{|c|c|c|c|c|c|}
\hline
 \, & $A_1$ & $A_2$ & $A_3$ & $q$ & $\tau$ \\
\hline   
 Mean & 7.66 & -2.18 & 34.31 & 0.557 & 0.549 \\
\hline    
 SD & 0.51 & 0.35 & 1.18 & 0.021 & 0.049 \\
\hline
\end{tabular}
\label{umean1}
\end{center}
\end{table} 

\begin{figure}[h!]
\centering
\includegraphics[width=22cm]{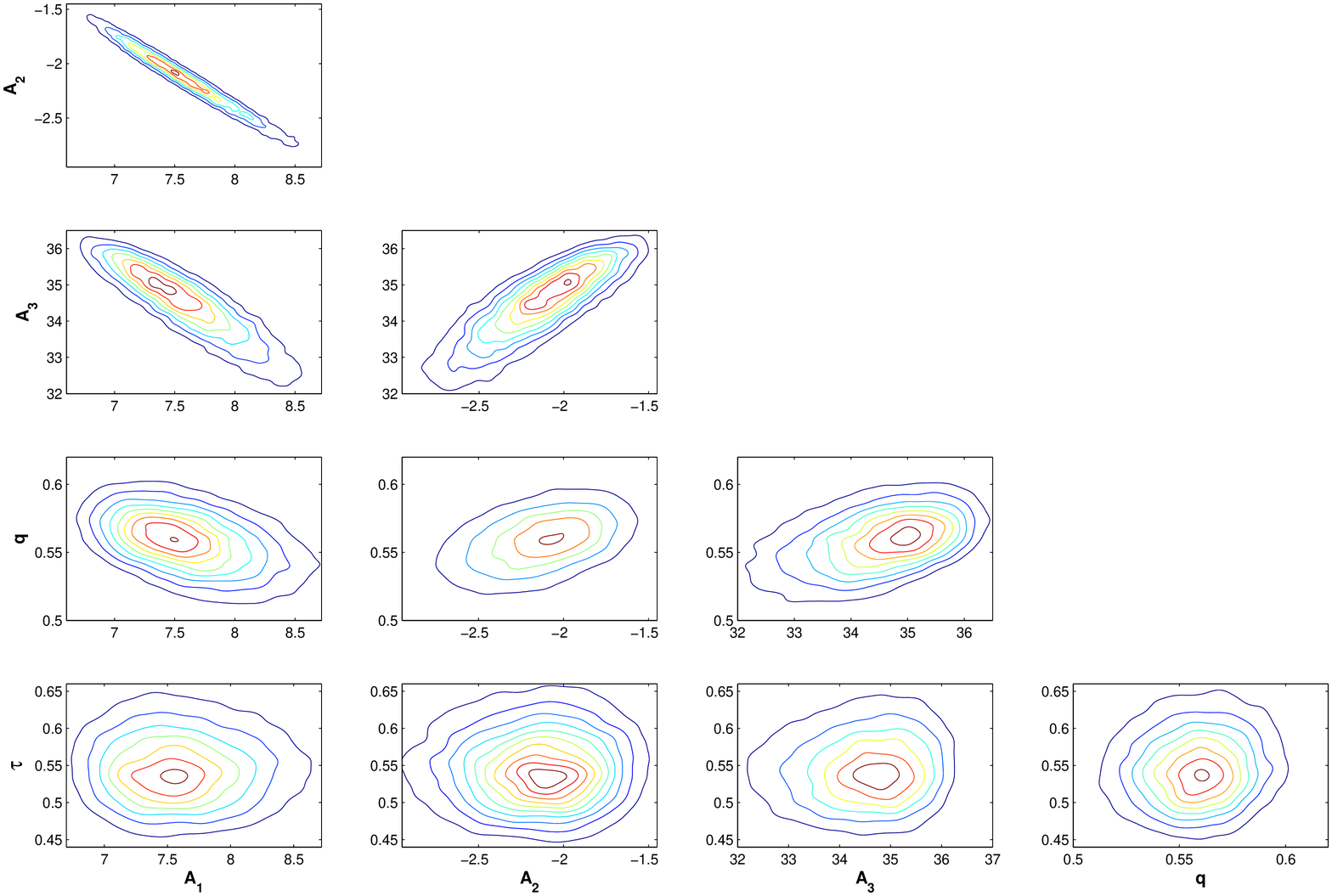}
\caption{Contour plots of the approximate bivariate densities of each pair of parameters in Model Ia. Similar to Figure \ref{biv1}, there is a strong correlation between $A_1$ and $A_2$. In addition, the fatigue limit parameter, $A_3$, is highly correlated with $A_1$ and $A_2$.}
\label{ubiv1}
\end{figure}

Maximum posterior estimates presented in Table \ref{uMAP1} are similar to the maximum posterior estimates obtained for Model Ia and reported in Table \ref{MAP1}. Table \ref{umean1} shows that the MCMC posterior samples obtained for Model Ia with uniform priors have more variability than the posterior samples obtained for Model Ia under Scenario 1. However, Figure \ref{uMCMC1} shows that the data are informative for all the parameters. The marginal posterior of the fatigue limit parameter is again left-skewed but with larger variance. Correlation coefficients given in Table \ref{uTbiv1} and Figure \ref{ubiv1} show that $A_1$ and $A_2$ are again linearly dependent and the number of parameters for Model Ia can be reduced by one parameter. In general, the correlations among the parameters of Model Ia under Scenario 2 have increased when compared with those under Scenario 1 (Table \ref{biv1}).

\begin{table}[h!]
\begin{center}
\caption{Correlation coefficients between each pair of parameters in Model Ia.} 
\begin{tabular}{|c|c|c|c|c|}
\hline
   & $A_1$ & $A_2$ & $A_3$ & $q$ \\
\hline   
$A_2$ & -0.986 & --- & --- & ---  \\
\hline    
$A_3$ & -0.908 & 0.863 &  ---  & ---  \\
\hline
 $q$ & -0.447 & 0.430 & 0.448 & ---  \\
\hline
$\tau$ & -0.017 & -0.018 & 0.018 & 0.060 \\
\hline
\end{tabular}
\label{uTbiv1}
\end{center}
\end{table}

\newpage \textbf{Numerical Results - Model Ib}

\begin{figure}[h!]
\centering
\includegraphics[width=17cm]{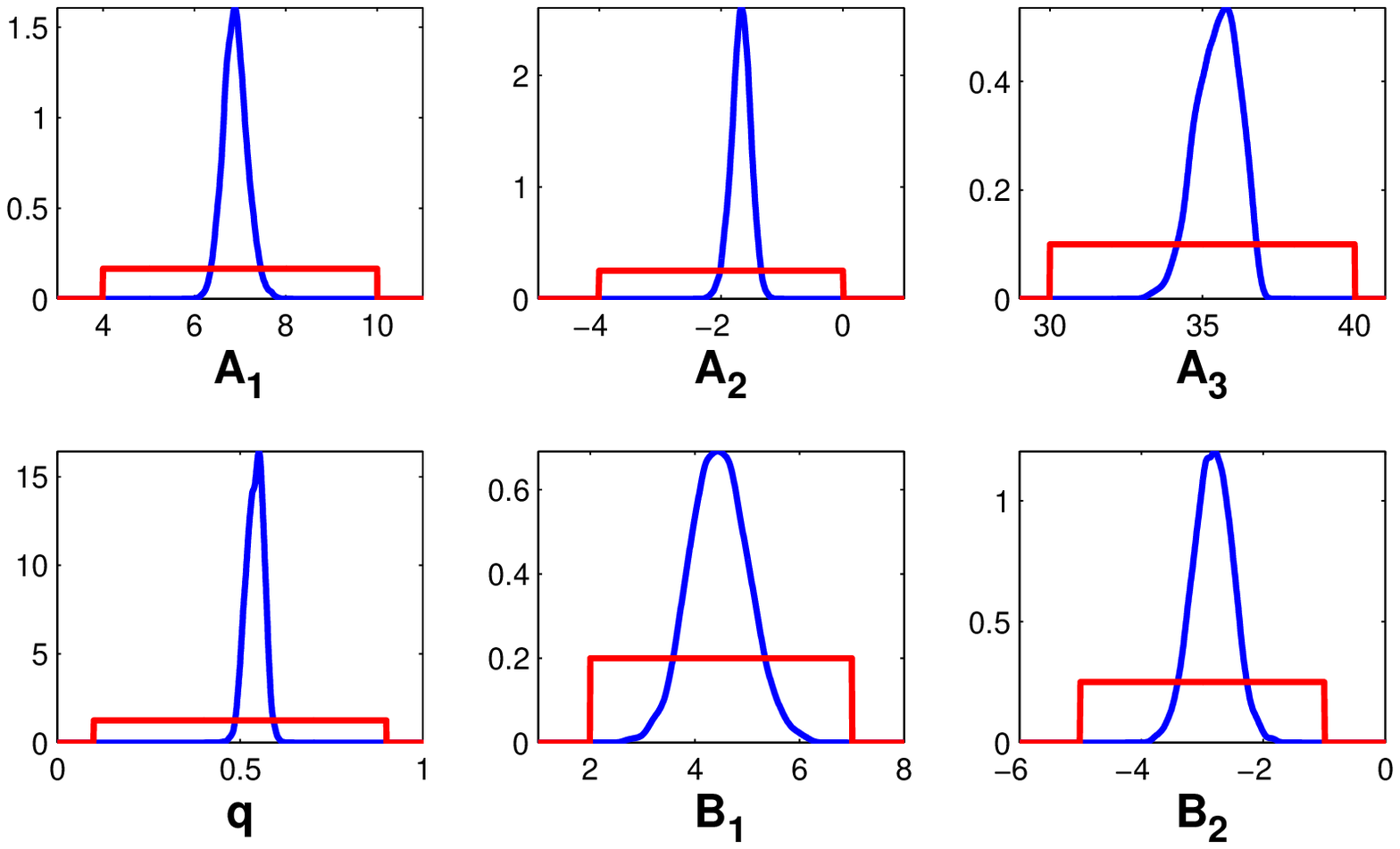}
\caption{Prior densities (red line) and approximate marginal posterior densities (blue line) for $A_1, A_2, q, B_1, B_2$ and $A_3\,.$ The estimated posterior densities for all parameters are concentrated around the corresponding modes but they show greater variability in comparison with Figure \ref{MCMC1b}.}
\label{uMCMC1b}
\end{figure}   

\begin{table}[h!]
\begin{center}
\caption{Maximum posterior estimates for Model Ib.}
\begin{tabular}{|c|c|c|c|c|c|c|}
\hline
Estimator & $A_1$ & $A_2$ & $A_3$ & $q$ & $B_1$ & $B_2$ \\
\hline
Laplace & 6.72 & -1.57 & 36.21 & 0.551 & 4.55 & -2.89 \\
\hline
Laplace-Metropolis & 6.69 & -1.55 & 36.19 & 0.551 & 4.76 & -3.00\\
\hline
\end{tabular}
\label{uMAP1b}
\end{center}
\end{table}

\begin{table}[h!]
\begin{center}
\caption{MCMC posterior empirical mean estimates with their standard deviations.} 
\begin{tabular}{|c|c|c|c|c|c|c|}
\hline
 \, & $A_1$ & $A_2$ & $A_3$ & $q$ & $B_1$ & $B_2$ \\
\hline   
 Mean & 6.90 & -1.67 & 35.50 & 0.541 & 4.46 & -2.83 \\
\hline    
 SD & 0.26 & 0.16 & 0.69 & 0.024 & 0.55 & 0.32 \\
\hline
\end{tabular}
\label{umean1b}
\end{center}
\end{table}

\begin{figure}[h!]
\centering
\includegraphics[width=22cm]{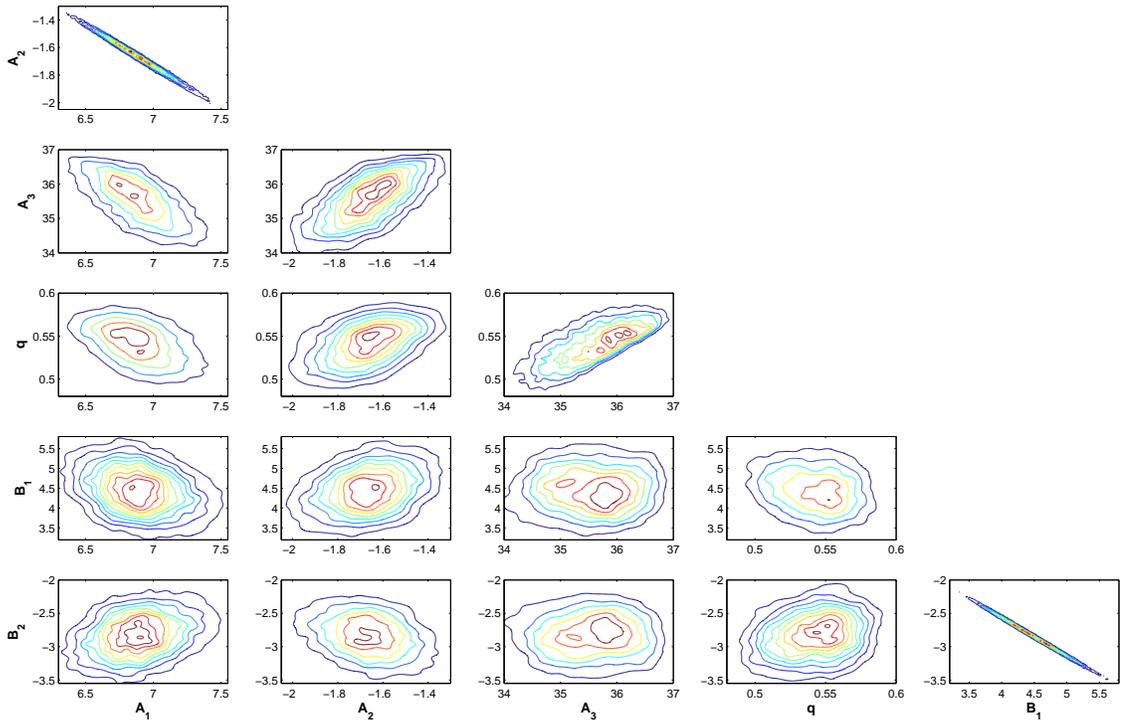}
\caption{Contour plots of the approximate bivariate densities of each pair of parameters in Model Ib. Similar to Figure \ref{biv1b}, there are two strong correlations between $A_1$ and $A_2$ and between $B_1$ and $B_2$. The fatigue limit parameter, $A_3$, has a moderate correlation with $A_1, A_2$ and $q$.  Allowing a non-constant variance has the effect of increasing the correlation between $q$ and the fatigue limit parameter, $A_3$.}
\label{ubiv1b}
\end{figure}

Maximum posterior estimates presented in Table \ref{uMAP1b} are similar to the maximum posterior estimates obtained for Model Ib and given in Table \ref{MAP1b}. Similar to Scenario 1, Figure \ref{uMCMC1b} and Table \ref{umean1b} show that the fatigue limit parameter, $A_3$, is again the most uncertain parameter whereas $q$ is the least uncertain parameter. However, the MCMC posterior samples obtained with uniform priors have more variability than do the posterior samples in Scenario 1 (Table \ref{mean1b}). The marginal posterior of the fatigue limit parameter is again left-skewed but with larger variance. Correlation coefficients in Table \ref{uTbiv1b} and Figure \ref{ubiv1b} show that $A_1$ and $B_1$ are almost perfectly correlated with $A_2$ and $B_2$, respectively. The correlations among the $A_1, A_2, A_3$ and $q$ parameters under Scenario 2 have increased when compared with those under Scenario 1 (Table \ref{biv1b}).

\begin{table}[h!]
\begin{center}
\caption{Correlation coefficients between each pair of parameters in Model Ib.} 
\begin{tabular}{|c|c|c|c|c|c|}
\hline
   & $A_1$ & $A_2$ & $A_3$ & $q$ & $B_1$ \\
\hline   
$A_2$ & -0.995 & --- & --- & --- & --- \\
\hline    
$A_3$ & -0.671 & 0.653 & --- & --- & --- \\
\hline
 $q$ & -0.428 & 0.436 & 0.664 & --- & --- \\
\hline
$B_1$ & -0.272 & 0.279 & 0.001 & -0.226 & ---\\
\hline
$B_2$ & 0.271 & -0.278 & 0.004 & 0.235 & -0.998 \\
\hline
\end{tabular}
\label{uTbiv1b}
\end{center}
\end{table}

\newpage \textbf{Numerical Results - Model IIb}

\begin{figure}[h!]
\centering
\includegraphics[width=17cm]{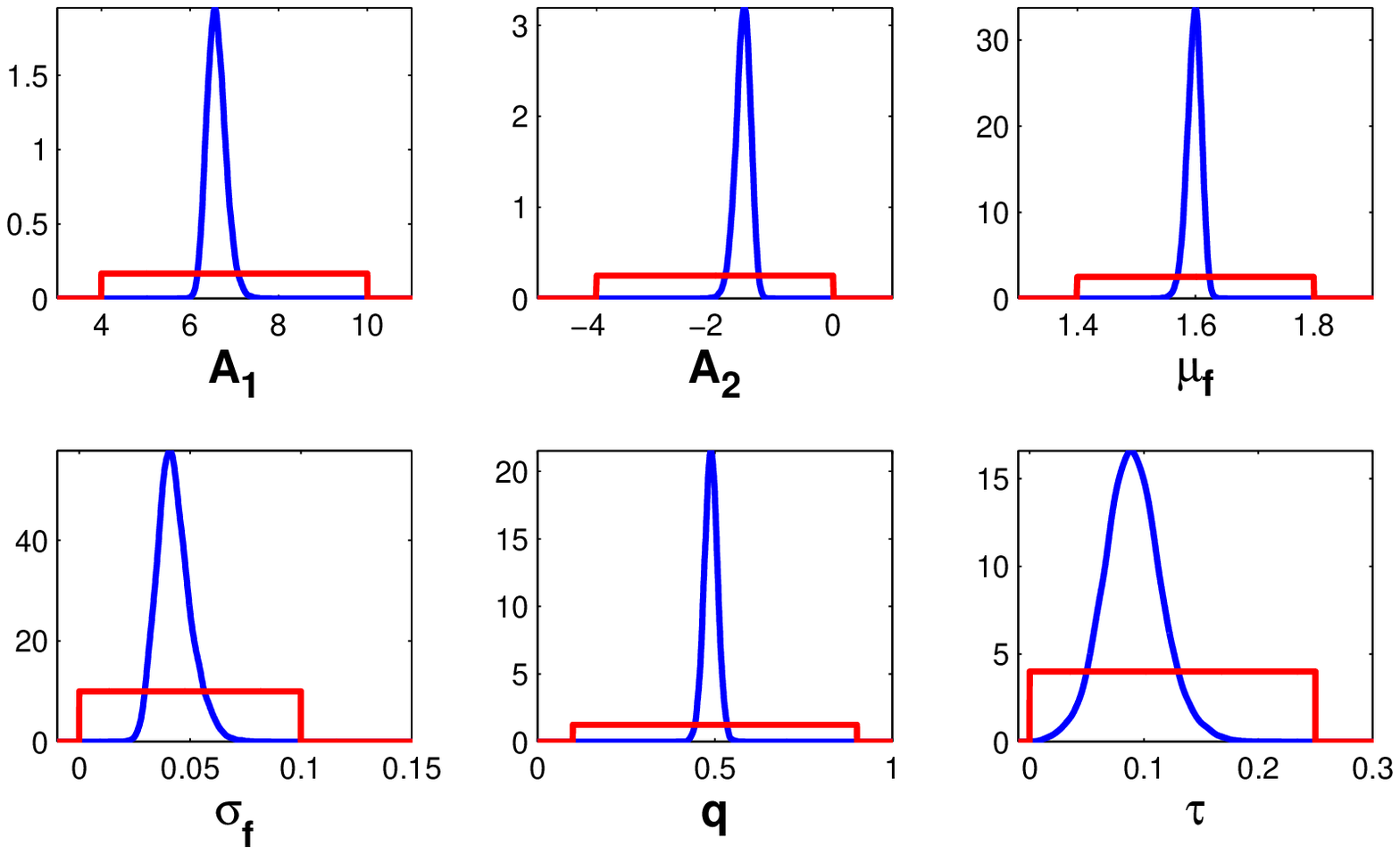}
\caption{Prior densities (red line) and approximate marginal posterior densities (blue line) for $A_1, A_2, q,\tau, \mu_{f}$ and $\sigma_{f}$. The estimated posterior densities for all parameters are concentrated around the corresponding modes but they show slightly greater variability in comparison with Figure \ref{MCMCFL}.}
\label{uMCMCFL}
\end{figure}   

\begin{table}[h!]
\begin{center}
\caption{Maximum posterior estimates for Model IIb.}
\begin{tabular}{|c|c|c|c|c|c|c|}
\hline
Estimator & $A_1$ & $A_2$ & $\mu_{f}$ & $\sigma_{f}$ & $q$ & $\tau$ \\
\hline
Laplace &  6.51 &  -1.48 &  1.60 &  0.0386 &  0.489 &   0.0853 \\
\hline
Laplace-Metropolis & 6.57 & -1.51 & 1.60 & 0.0398 & 0.490 & 0.0856 \\
\hline
\end{tabular}
\label{uMAPFL}
\end{center}
\end{table}

\begin{table}[h!]
\begin{center}
\caption{MCMC posterior empirical mean estimates with their standard deviations.} 
\begin{tabular}{|c|c|c|c|c|c|c|}
\hline
 \, & $A_1$ & $A_2$ & $\mu_{f}$ & $\sigma_{f}$ & $q$ & $\tau$ \\
\hline   
 Mean & 6.60 &  -1.52 &  1.60 &  0.0428 &  0.489 &   0.0901 \\
\hline    
 SD & 0.21 &  0.13 & 0.012 &  0.008 &  0.019 & 0.025 \\
\hline
\end{tabular}
\label{umeanFL}
\end{center}
\end{table} 

\begin{figure}[h!]
\centering
\includegraphics[width=22cm]{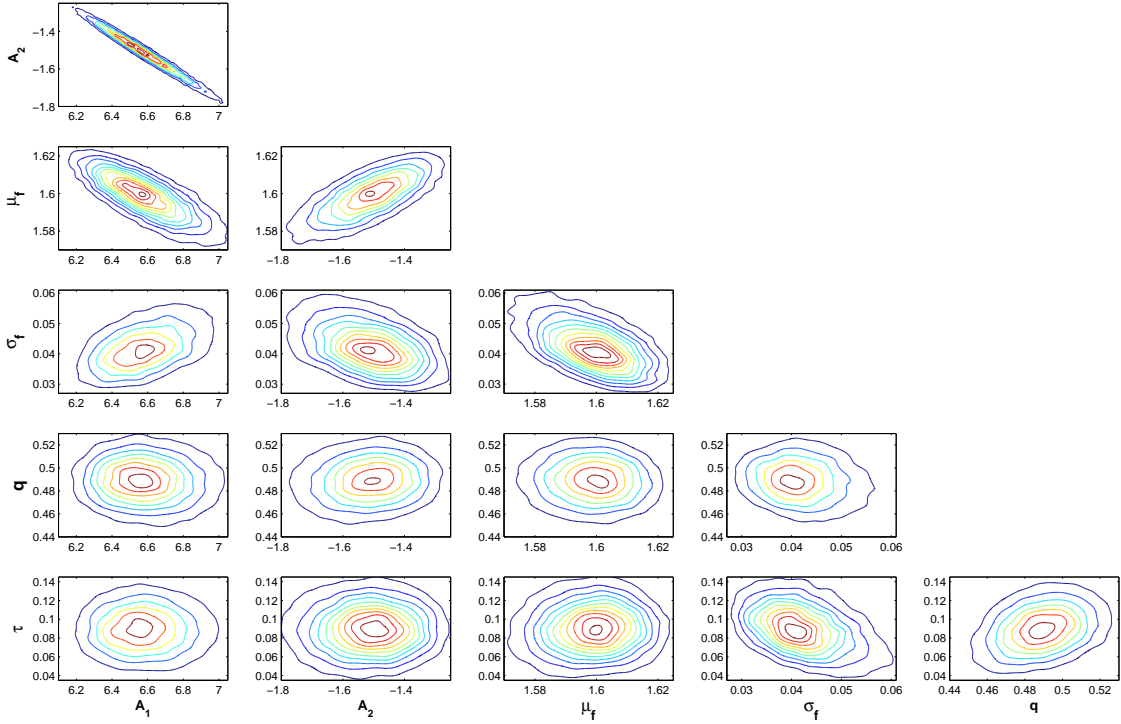}
\caption{Contour plots of the approximate bivariate densities of each pair of parameters in Model IIb. Similar to Figure \ref{bivFL}, the parameter $\mu_f$ shows a strong correlation with $A_1$ and $A_2$. The random fatigue-limit model has the effect of reducing the correlations between $q$ and the parameters $A_1$ and $A_2$.}
\label{ubivFL}
\end{figure}

Maximum posterior estimates given in Table \ref{uMAPFL} are similar to the maximum posterior estimates obtained for Model IIb under Scenario 1 and presented in Table \ref{MAPFL}. Similar to Scenario 1, Figure \ref{uMCMCFL} and Table \ref{umeanFL} show that the location and scale parameters, $\mu_f$ and $\sigma_f$, are the least uncertain parameters. The MCMC posterior samples obtained with uniform priors have slightly greater variability than do the posterior samples in Scenario 1 (Table \ref{meanFL}). Figure \ref{uMCMCFL} shows that the data are very informative for all the parameters because there is a strong contraction between the prior densities and the posterior densities. Table \ref{uTbivFL} and Figure \ref{ubivFL} show that $A_1$ and $A_2$ can be considered linear dependent, and therefore we can reduce the number of parameters for Model IIb by one parameter. In general, the correlations among the parameters of Model IIb under Scenario 2 are greater than those correlations under Scenario 1 (Table \ref{bivFL}).

\begin{table}[h!]
\begin{center}
\caption{Correlation coefficients between each pair of parameters in Model IIb.} 
\begin{tabular}{|c|c|c|c|c|c|}
\hline
  & $A_1$ & $A_2$ & $\mu_{f}$ & $\sigma_{f}$ & $q$ \\
\hline   
$A_2$ & -0.987 & --- & --- & --- & --- \\
\hline    
$\mu_{f}$  & -0.788 & 0.722 &  ---  & ---  & --- \\
\hline
$\sigma_{f}$ & 0.494 & -0.452 & -0.566 & --- & --- \\
\hline
$q$ &  -0.067 & 0.111 & -0.037 & -0.158 & --- \\
\hline
$\tau$ &  -0.019  & 0.031 & 0.080 & -0.412 & 0.326 \\
\hline
\end{tabular}
\label{uTbivFL}
\end{center}
\end{table} 

\subsection{Model comparison} \label{sec4b}
We now analyze more closely comparisons among models  Ia, Ib and  IIb.

\subsubsection{Bayes Factor}
\label{BF}
We adopt a traditional Bayesian approach by estimating the Bayes factor of Model A against that of Model B, which is defined as

\begin{equation*}
F_{B\,,A} : = \frac{\int L_{B}(\theta_{B} ; \textbf{y}) \rho_{B}(\theta_{B}) d{\theta_{B}}}{\int L_{A}(\theta_{A} ; \textbf{y}) \rho_{A}(\theta_{A}) d{\theta_{A}}} = \frac{ p_{B}(\textbf{y})}{ p_{A}(\textbf{y})} \,,
\end{equation*}
where $\rho_{A}(\mathbf{\theta}_{A})$ and $\rho_{B}(\mathbf{\theta}_{B})$ are the prior densities, and $p_{A}(\textbf{y})$ and  $p_{A}(\textbf{y})$ are the marginal likelihoods \cite[Chapter 2]{con}.

Common methods to estimate Bayes factors \cite{dici, lewraf} are applied to compare the fitted models and to rank their plausibility. Fast preliminary estimates of the log marginal likelihoods were obtained through the application of the Laplace approximation. Then, the log marginal likelihoods were computed using the Laplace-Metropolis estimator, which is based on the MCMC posterior samples together with the Laplace approximation. 

In both cases, the approximation of the log marginal likelihood $\log(p(\textbf{y}))$ is given by

\begin{equation*}
\frac{P}{2} \log(2 \pi) + \frac{1}{2} \log(|H^*|) +  \log\left( \rho(\theta^*) \right) + \log\left( L(\theta^* | \textbf{y}) \right),
\end{equation*}
where $P$ is the dimension of the vector $\theta$, $\theta^*$ is the maximum posterior estimate and $H^*$ is the inverse Hessian of the negative log posterior.

In the Laplace estimator, $\theta^*$ and $H^*$ are numerically approximated by means of the Broyden-Fletcher-Goldfarb-Shanno (BFGS) algorithm. The Laplace-Metropolis estimator uses the MCMC posterior samples to find the maximum posterior estimate, $\theta^*$, and approximate, $H^*$, by the empirical covariance matrix.

\subsubsection{Predictive Information Criteria for Bayesian Models}
\label{PIC}
In this section, we compare models by measuring their prediction accuracy. We estimate the prediction accuracy using deviance and Watanabe-Akaike information criteria as well as cross-validation. 

%Let $\{ y_i \}_{i=1}^{n}$ be the dataset observations and $\{ \theta^m \}_{m=1}^{S}$ be the MCMC posterior samples.

\begin{itemize}

\item \textbf{Log pointwise predictive density (lppd)} \\
The general method to estimate the prediction accuracy of a certain model is through the log predictive density, $\log \rho(y|\theta) = \log L(\theta; y)$, where $y$ is a new observation. An overestimate of the log predictive density can be obtained by using the observed data, $\{ y_i \}_{i=1}^{n}$. It is an overestimate because the observed data were used first to infer $\theta$. In our Bayesian approach, $\theta$ is summarized by the MCMC posterior samples, $\{ \theta^m \}_{m=1}^{S}$, and therefore the log pointwise predictive density estimate is given by

\begin{equation}
\label{lppd}
lppd = \sum_{i=1}^{n} \log \left( \frac{1}{S} \sum_{m=1}^{S} \rho(y_{i} | \theta^m) \right),
\end{equation}
where $S$ should be ``large enough" \cite{gelhv, gelv}.

\item \textbf{Deviance information criterion (DIC)}\\
DIC can be considered as a Bayesian generalization of the AIC by replacing the maximum likelihood estimate by the posterior mean and computing the effective number of parameters, $p_{DIC}$, as follows:
\begin{equation*}
p_{\text{DIC}} = 2 \left( \log L(\bar{\theta}) - \frac{1}{S} \sum_{m=1}^{S} \log L(\theta^m) \right),
\end{equation*}
where $\bar{\theta}$ is the posterior mean \cite{gelhv}. Then, the deviance information criterion is given by
\begin{equation*}
\text{DIC} = -2 \left( \log L(\bar{\theta}) - p_{DIC} \right).
\end{equation*}

\item \textbf{Watanabe-Akaike information criterion (WAIC)}\\
WAIC or widely applicable information criterion is a stable Bayesian predictive measure that approximates the leave-one-out cross-validation (see \cite{gelhv, gelv, waic}) and is defined by
\begin{equation*}
p_{\text{WAIC}} = 2\sum_{i=1}^{n} \left( \log \left( \frac{1}{S} \sum_{m=1}^{S} \rho(y_{i} | \theta^m) \right) - \frac{1}{S} \sum_{m=1}^{S} \log \rho(y_{i} | \theta^m) \right),
\end{equation*}
\begin{equation*}
\text{WAIC} = -2(lppd - p_{\text{WAIC}}).
\end{equation*}

\item \textbf{K-fold cross-validation} \\
Cross-validation is the most popular yet computationally expensive method to estimate a model's predictive accuracy. We consider the K-fold cross-validation where the data are randomly partitioned into K disjoint subsets, $\{\bold{y}_k\}_{k = 1}^{K}$. Then, we define $\{\bold{y}_{(-k)}\}= \{\bold{y}_1, \ldots, \bold{y}_{k-1}, \bold{y}_{k+1}, \dots, \bold{y}_K\}$ to be a training set. For each training set, we compute the corresponding posterior distribution, $p(\theta | \bold{y}_{(-k)})$. Then, the log predictive density for $y_i \in \bold{y}_{k}$ is computed using the training set $\{\bold{y}_{(-k)}\}$, that is:
\begin{equation*}
lpd_{i} = \log \left( \frac{1}{S} \sum_{m=1}^{S} \rho(y_{i} | \theta^{k,m}) \right), i \in k,
\end{equation*}
where $\{ \theta^{k,m} \}_{m=1}^{S}$ are the MCMC samples of the posterior $p(\theta|\bold{y}_{(-k)})$. Finally, we sum to obtain the expected log predictive density (elpd):
\begin{equation*}
elpd = \sum_{i=1}^{n} lpd_{i}.
\end{equation*}

The K-fold cross-validation (with $K=5$ or $10$) is usually used instead of the leave-one-out cross-validation, which is the most computationally exhaustive type of cross-validation (see \cite[Chapter 5]{Izenman} and \cite{gelv}).
\end{itemize}

In the next Subsection, we present the main numerical results from applying the techniques described in Subsections \ref{BF} and \ref{PIC} for Models Ia, Ib and IIb under the two predefined scenarios. 

\subsubsection{Numerical Results (Scenario 1)}

\begin{table}[h!]
\begin{center}
\caption{Log marginal likelihoods (Bayes factors) show very strong evidence that Model Ib is better than Model Ia and that Model IIb is better than Model Ib. The predictive information criteria and the 5-fold cross-validation show that Model IIb also has better predictive accuracy than do Model Ia and Model Ib.}
\begin{tabular}{|c|c|c|c|c|}
\hline
Models & \bf{Model Ia} & \bf{Model Ib} & \bf{Model IIb}\\
\hline
Log marginal likelihood  (Laplace) & -963.07 & -940.18 & -932.55 \\
\hline
Log marginal likelihood (Laplace-Metropolis) & -963.16  & -937.06 & -923.68 \\
\hline
Log pointwise predictive density (lppd) & -949.56 & -920.51 & -907.85 \\
\hline
Deviance information criterion (DIC) & 1909.6 & 1851.8 & 1826.5 \\
\hline
Watanabe-Akaike information criterion (WAIC) & 1911.3 & 1853.1 & 1825.9\\
\hline
5-fold cross-validation elpd & -955.42 & -927.07  & -913.80 \\
\hline
%Effective number of parameters & 5.8643 & 6.5643 &  5.9548 \\
%\hline
\end{tabular}
\label{Bapp1}
\end{center}
\end{table}

Table \ref{Bapp1} shows that Model IIb under Scenario 1 is preferable by the log marginal likelihood and the predictive information criteria. The Laplace method appears to underestimate the log marginal likelihood for Model IIb. This is expected because of the complex likelihood function of Model IIb and because the Gaussian approximation does not always provide a good estimation. Table \ref{Bapp1} also shows consistency with the classical information criterion presented in Table \ref{Capp}.

\subsubsection{Numerical Results (Scenario 2)}

Table \ref{Bapp2} shows that Model IIb under Scenario 2 is preferable by the log marginal likelihood and the predictive information criteria. In general, the estimated values in Table \ref{Bapp2} are slightly higher in magnitude than are the results in Table \ref{Bapp1}. This is reasonable because Bayesian models with proper informative priors should be preferable. 

\begin{table}[h!]
\begin{center}
\caption{Log marginal likelihoods (Bayes factors) indicate that Model Ib is better than Model Ia and that Model IIb is better than Model Ib. The predictive information criteria and the 5-fold cross-validation show that Model IIb also has better predictive accuracy than do Model Ia and Model Ib.}
\begin{tabular}{|c|c|c|c|c|}
\hline
Models & \bf{Model Ia} & \bf{Model Ib} & \bf{Model IIb}\\
\hline
Log marginal likelihood  (Laplace) & -963.36 & -940.25 & -923.91 \\
\hline
Log marginal likelihood (Laplace-Metropolis) & -963.51  & -938.17 & -923.76 \\
\hline
Log pointwise predictive density (lppd) & -949.70 & -920.52 & -908.01 \\
\hline
Deviance information criterion (DIC) & 1910.4 & 1852.4 & 1827.1 \\
\hline
Watanabe-Akaike information criterion (WAIC) & 1912.1 & 1853.9 & 1826.4 \\
\hline
5-fold cross-validation elpd & -955.70 & -927.80 & -914.07 \\
\hline
%Effective number of parameters & 5.9979 & 7.2745  & 6.0627 \\
%\hline
\end{tabular}
\label{Bapp2}
\end{center}
\end{table}

\section{Conclusions}

We calibrated models of various complexity that were designed to account for right-censored data by means of the maximum likelihood method. We used a data set described in Section 2 for this purpose. The robustness of the estimation of the quantile functions has been assessed by computing bootstrap confidence intervals for samples stratified with respect to the cycle ratio. 

We then considerably enlarged the scope of our study by considering a Bayesian approach. Any prior distribution, which is suitable to describe the available knowledge on the physical parameters, can be easily incorporated into our Bayesian computational framework that provides a simulation-based posterior distribution.  

To decide which model could be considered more reliable for deployment, first we computed classical measures of fit based on information criteria. Then, the Bayesian approach for model comparison was applied to determine which model would be preferred under different a priori scenarios. This approach included very different techniques ranging from those based on the estimation of the marginal likelihood to those based on predictive information criteria, whose implementation require the use of cross-validation techniques. 

The classical approach and the Bayesian approach for model comparison have provided evidence in favor of Model IIb given the 75S-T6 data set described in Section 2. Model IIb assumes that both fatigue life and the fatigue limit parameter follow a Weibull distribution and the expected value of the fatigue limit parameter, $A_3$, is $39.88$ ksi. 

An integrated set of computational tools has been developed for model calibration, cross-validation, consistency and model comparison, allowing the user to rank alternative statistical models based on objective criteria.

\section*{Acknowledgement}
Z. Sawlan, M. Scavino and R. Tempone are members of King Abdullah University of Science and Technology (KAUST) SRI Center for Uncertainty Quantification in Computational Science and Engineering.

%The research reported in this publication was supported by King Abdullah University of Science and Technology (KAUST) through support from the Strategic Research Initiative on Uncertainty Quantification to Z. Sawlan, M. Scavino and R. Tempone.

%\section*{References}

\bibliography{mybib}

\end{document}